\documentclass[groupedaddress,nofootinbib,11pt,preprintnumbers]{revtex4-1}
\usepackage[utf8]{inputenc}

\usepackage[dvipsnames]{xcolor}
\definecolor{red}{rgb}{0.9, 0,0}
\definecolor{cerulean}{rgb}{0., 0.42,0.9}
\definecolor{navy}{rgb}{0.05, 0.05,0.8}

\usepackage[colorlinks]{hyperref}
\hypersetup{
    colorlinks = true,
    citecolor  = red,
	linkcolor  = navy
}

\usepackage{slashed}
\usepackage{graphics}
\usepackage{amsmath}
\usepackage{amssymb}
\usepackage{latexsym}
\usepackage{mathtools}
\usepackage{dsfont}
\usepackage{hyperref}
\usepackage{amsfonts}
\usepackage{enumitem}
\usepackage{xstring} 
\usepackage{xspace} 
\usepackage[normalem]{ulem}
\usepackage{fontawesome}
\usepackage{subcaption}

\newcommand{\ns}{\,{\rm ns}}

\newcommand{\eV}{\,{\rm eV}}

\newcommand{\MeV}{\,{\rm MeV}}
\newcommand{\GeV}{\,{\rm GeV}}

\newcommand{\pc}{\,{\rm pc}}
\newcommand{\kpc}{\,{\rm kpc}}
\newcommand{\Mpc}{\,{\rm Mpc}}
\newcommand{\wk}{\,{\rm weeks}}
\newcommand{\yr}{\,{\rm years}}

\def\vec#1{\mathbf{#1}}

\newcommand{\eq}[1]{Eq.~\eqref{#1}}
\newcommand{\Fig}[1]{Fig.~\ref{#1}}

\newcommand{\githublink}{\href{https://github.com/szehiml/dm-pta-mc}{\faGithub}\xspace}

 \definecolor{lightblue}{rgb}{0.2,0.5,1}
 \hypersetup{colorlinks=true,
 linkcolor=lightblue,
 citecolor=lightblue,
 urlcolor=lightblue,
 linktocpage=lightblue,
 pdfproducer=lightblue}
\begin{document}

\title{Probing Small-Scale Power Spectra with Pulsar Timing Arrays}

\author{Vincent S. H. Lee}
\author{Andrea Mitridate} 
\author{Tanner Trickle}
\author{Kathryn M. Zurek}
\affiliation{Walter Burke Institute for Theoretical Physics, California Institute of Technology, Pasadena, CA}

\preprint{CALT-TH-2020-058}

\begin{abstract}
Models of Dark Matter (DM) can leave unique imprints on the Universe's small scale structure by boosting density perturbations on small scales.  We study the capability of Pulsar Timing Arrays to search for, and constrain, subhalos from such models. The models of DM we consider are ordinary adiabatic perturbations in $\Lambda$CDM, QCD axion miniclusters, models with early matter domination, and vector DM produced during inflation. We show that $\Lambda$CDM, largely due to tidal stripping effects in the Milky Way, is out of reach for PTAs (as well as every other probe proposed to detect DM small scale structure). Axion miniclusters may be within reach, although this depends crucially on whether the axion relic density is dominated by the misalignment or string contribution.  Models where there is matter domination with a reheat temperature below 1 GeV may be observed with future PTAs.  Lastly, vector DM produced during inflation can be detected if it is lighter than $10^{-16} \mbox{ GeV}$.  We also make publicly available a Python Monte Carlo tool for generating the PTA time delay signal from any model of DM substructure.
\end{abstract}


\maketitle
\newpage
\tableofcontents
\newpage

\section{Introduction}

Dark Matter (DM) is one of the pillars of the standard cosmological model. Perturbations in the DM density field generated by inflation provide the seeds for the hierarchical structure formation we observe in the Universe. On large scales the matter power spectrum of these primordial perturbations can be inferred from the anisotropies in the cosmic microwave background (CMB). These observations indicate a nearly scale-invariant spectrum of primordial fluctuations, which is compatible with the large scale structures we observe on galactic and extra-galactic scales.

However on smaller length scales, $k\gtrsim\pc^{-1}$, different theories of DM leave unique fingerprints on the primordial perturbations and/or their evolution. Here we focus on four theories which produce different small scale structures, avoid current experimental bounds, and are theoretically well motivated:

\begin{itemize}
    \item \textbf{$\Lambda$CDM}: A nearly scale invariant spectrum of adiabatic perturbations is produced at the end of infaltion~\cite{Kolb1994, Dodelson2003}.
    \item \textbf{Post-inflationary QCD axion}: The $U(1)_\text{PQ}$ symmetry is broken after inflation. The decay of axion field defects at the QCD phase transition then induces large amplitude isocurvature fluctuations on scales smaller than the horizon at the QCD epoch~\cite{Hogan:1988mp,Kolb:1993zz}.
    \item \textbf{Early Matter Domination}:  Adiabatic perturbations which are within the horizon during an early stage of matter domination grow linearly instead of logarithmically, enhancing the amount structure on these scales~\cite{Erickcek:2011us, Fan:2014zua}.
    \item \textbf{Vector DM Produced During Inflation}: If the DM is a massive spin-1 particle, the longitudinal modes produced at the end of inflation can peak the power spectrum at small scales, with the location of the peak determined by the DM mass~\cite{Graham:2015rva}.
\end{itemize}
These model specific features in the primordial seeds, and their evolution, translate to different predictions for the amount, and properties, of sub-galactic DM halos (subhalos). Therefore measuring this population of subhalos can be a powerful tool in pinning down the model of DM. 

Unfortunately DM subhalos are elusive objects, mostly because they are expected to contain very little baryonic matter and are therefore nearly invisible (assuming a weak indirect detection signal). Because of this, gravitational probes are the natural candidate to look for them. Many such probes have been proposed (\emph{e.g.} Refs.~\cite{VanTilburg:2018ykj, Rivero:2017mao, Metcalf:2001ap, Dai:2019lud}), however their discovery potential depends on the subhalo mass and density profile (usually parameterized with a concentration parameter). At small masses ($M\lesssim10^{-2}M_\odot$), where these probes lose sensitivity, Pulsar Timing Arrays (PTAs) may be powerful and complementary probes  \cite{Siegel:2007fz, Baghram:2011is, Clark:2015sha, Schutz:2016khr, Dror:2019twh, Ramani:2020hdo}. The ability to test extremely light ($M \gtrsim10^{-13}$) subhalos with low concentration parameters ($c \gtrsim 10$) makes PTA searches particularly interesting.

The signals DM subhalos can induce in a PTA measurement of the pulsar phase have been studied in depth~\cite{Dror:2019twh, Ramani:2020hdo}. Each subhalo induces a shift to the residual phase measurement which can be non-degenerate with the timing model. Previous works \cite{Dror:2019twh, Ramani:2020hdo} have used analytic approximations of the signal, denoted as \textit{static}, \textit{dynamic}, and \textit{stochastic} limits.  These correspond to three different limits where the timescale, $\tau$, for a typical subhalo to transit the line-of-sight is much greater than the observing time $T$ ($\tau \gg T$, static), is much less than the observing time ($\tau \ll T$), but the signal is dominated by a single transiting subhalo (dynamic) or many transiting subhalos (stochastic). Here we will limit these simplifications by generating the full time series produced by a population of DM halos by using a Monte Carlo (MC) simulation. One goal of this paper is to develop this numeric tool, which we make publicly available on GitHub \githublink.

The second goal of this work is to use this MC to generate the PTA signal that would be produced by the subhalo populations predicted by the four DM models listed above. To obtain an estimate of the local population of DM subhalos, one needs to know how primordial perturbations evolve from the early Universe until today. This is a complex problem, especially at the sub-galactic scales relevant for PTA searches where non-linearities and tidal effects play a crucial role. The Press-Schechter formalism \cite{Press:1973iz} is known to give a reasonably good analytic description of the non-linear physics related to the clustering of DM overdensities (at least for the case of $\Lambda$CDM, where a direct comparison with numerical simulations is possible). We then use this model, together with semi-analytic description of tidal effects, to relate the primordial power of primordial perturbations to the local population of DM subhalos. We should caution that a number of the methods we are using have not be tested against $N$-body simulations for such low mass and high density subhalos. We leave for future work validating and calibrating the analytic results for small scale DM subhalos.

The outline of the paper is as follows. In Section \ref{sec:signal_deriv} we review the derivation of the PTA signal induced by a population of DM subhalos, along with the signal-to-noise ratio, and discuss the MC algorithm used to compute it. In Section \ref{sec:primordial_to_local} we illustrate the semi-analytic procedure used to relate the primordial power of density perturbation to the local subhalo population. In Section \ref{sec:constraints} we apply these results to the benchmark scenarios discussed above, and derive constraints for planned~\cite{Rosado:2015epa}, and future PTAs. Finally, in Section \ref{sec:conclusions} we conclude. 

\section{Dark Matter Subhalo Signatures in Pulsar Timing Arrays}
\label{sec:signal_deriv}

We begin with a discussion of the signal induced in PTAs by DM subhalos, how we generate this signal with the Monte Carlo, and the signal-to-noise ratio (SNR) of such a signal. Many of the formulae presented here were previously derived in Refs.~\cite{Dror:2019twh, Ramani:2020hdo} and we review them here for completeness. Previously both the Doppler effect (acceleration induced by a DM subhalo of either the Earth or a pulsar (called the {\em Earth} or \textit{pulsar} terms, respectively)), and Shapiro effect (a change in the light arrival time due to the DM subhalo gravitational potential) were considered. Here we will only consider signals from the Doppler effect, which is dominant over the Shapiro effect for subhalos with mass below $M \lesssim 10^{-3} M_\odot$ for any concentration parameter (see Ref.~\cite{Ramani:2020hdo}). 

\subsection{Phase Shifts from Dark Matter Subhalos}
Pulsars are rotating, highly magnetized neutron stars that emit beams of electromagnetic radiation from their magnetic poles. Given a misalignment between the rotation and magnetic axes, the pulsar rotation can cause the radiation beam to sweep across Earth. If this happens, a pulsar will appear to an Earth observer as a periodic emitter. When a DM subhalo approaches either the Earth, or a pulsar, in the array it will cause an acceleration and therefore change the observed pulsar frequency, $\nu$. This frequency shift, $\delta \nu$, for a subhalo passing with position $\vec{r} = \vec{r}_0 + \vec{v}t$, where $\vec{r}_0$ is the initial position and $\vec{v}$ is the velocity, is given by
\begin{align}
    \delta \nu \left( t; \vec{r}_0, \vec{v} \right) = \nu \, \hat{\vec{d}} \cdot \int^t_0 \nabla \Phi(\vec{r}(t')) \, dt' \, ,
    \label{eq:dnu_pot}
\end{align}
where $\Phi$ is the subhalo gravitational potential. The coordinate system for the Earth (pulsar) term is chosen with the Earth (pulsar) at the origin, with $\hat{\vec{d}}$ pointing from the Earth to the pulsar (pulsar to Earth). We also parameterize the position vector as $\vec{r}(t) = \vec{b} + \vec{v}(t - \overline{t})$, where $\vec{b} = \vec{r}_0 + \vec{v} \overline{t}$ is the impact parameter, and $\overline{t} = - \vec{r}_0 \cdot \vec{v} / v^2$ is the time to reach the point of closest approach. The phase shift, $\delta \phi$, is then simply
\begin{align}
    \delta \phi(t) = \int^t_0 \delta \nu(t') \, dt' \, .
    \label{eq:single_phase_shift}
\end{align}
For spherically symmetric halos, the gradient of gravitational potential appearing in \eq{eq:dnu_pot} can be written in terms of a form factor $\mathcal{F}(s,c)$ as
\begin{equation}\label{eq:halo_pot}
    \nabla \Phi(\vec{r},M) = \frac{GM}{r^3} \mathcal{F}(r/r_v, c) \,  \vec{r},
\end{equation}
where $M$ is the virial mass defined as the subhalo mass contained inside the virial radius $r_v$, the radius within which the mean halo density is 200 times the critical density of the Universe, $\rho_c$. In the following we will assume that DM subhalos have an NFW density profile
\begin{equation}
\rho(r) = \frac{4\rho_s}{\left( r/r_s \right) \left( 1 + r/r_s \right)^2}, \label{eq:nfw_profile}
\end{equation}
where $r_s$ is the scale radius, and $\rho_s \equiv \rho(r_s)$. For NFW subhalos the form factor is
\begin{equation}
\mathcal{F}(s, c) = \frac{\ln(1+c\,s)-c\,s/(1+c\,s)}{\ln(1+c)-c/(1+c)}, \label{eq:phi_form}
\end{equation}
where we have introduced the concentration parameter, $c\equiv r_v/r_s$, related to the subhalo scale density by 
\begin{equation}\label{eq:rho_to_c}
\rho_s = \frac{50\,c^3\rho_c}{3 \left( \ln(1 + c) - c/(1 + c) \right)}\,.
\end{equation}
We see that for very compact halos, $c \rightarrow \infty$, the form factor reduces to one and subhalos behave as point-like object. We will refer to this as the `PBH' limit since when $\mathcal{F} \rightarrow 1$ the gravitational potential reduces to that of a primordial black hole (PBH). 

In the PBH limit Eq.~\eqref{eq:single_phase_shift} can be simplified to \cite{Dror:2019twh}\footnote{Indefinite integrals were used in deriving Eq.~\eqref{eq:delta_phi_pbh} from Eq.~\eqref{eq:dnu_pot}. This will not affect any observable since any dependence on reference times will be removed in the fit to the residual phase shift.}
\begin{align}
    \delta \phi_\text{PBH}(t; M, \vec{r}_0, \vec{v}) = \frac{G M \nu }{v^2} \hat{\vec{d}} \cdot \left( \sqrt{1 + x^2} \,  \hat{\vec{b}} - \sinh^{-1}(x) \, \hat{\vec{v}} \right), 
    \label{eq:delta_phi_pbh}
\end{align}
where $x = (t - \overline{t})/\tau$ is a normalized time variable, and $\tau = b/v$.

For an $\mathcal{F}$ given by Eq.~\eqref{eq:phi_form}, $\delta \phi$ can be challenging to compute analytically due to the discontinous derivatve in $\mathcal{F}$ at $s = 1$ which can be important if the subhalo passes from  $s > 1$ to $s < 1$. In principle, one could perform the integrations numerically, though this becomes intensive with a large number of subhalos. We instead use a conservative estimate of the signal size by substituting $\mathcal{F}(r/r_v, c)$ with $\mathcal{F}^*(c) \equiv \mathcal{F}(r_\text{min}/r_v, c)$, where $r_\text{min}$ is the distance of closest approach over the observation time. This approximation changes the amplitude by $\mathcal{O}(10\%)$ for $c \sim 100$ but quickly shrinks to $\mathcal{O}(1\%)$ for $c \sim 10^4$. Using this conservative estimate the phase shift for an NFW subhalo is simply
\begin{align}
    \delta \phi(t; M, c, \vec{r}_0, \vec{v}) = \frac{G M \mathcal{F}^*(c) \nu}{v^2} \, \hat{\vec{d}} \cdot \left( \sqrt{1 + x^2} \,  \hat{\vec{b}} - \sinh^{-1}(x) \, \hat{\vec{v}} \right) \, .
    \label{eq:delta_phi_form}
\end{align}

\begin{figure}[t]
    \centering
    \includegraphics[width=\textwidth]{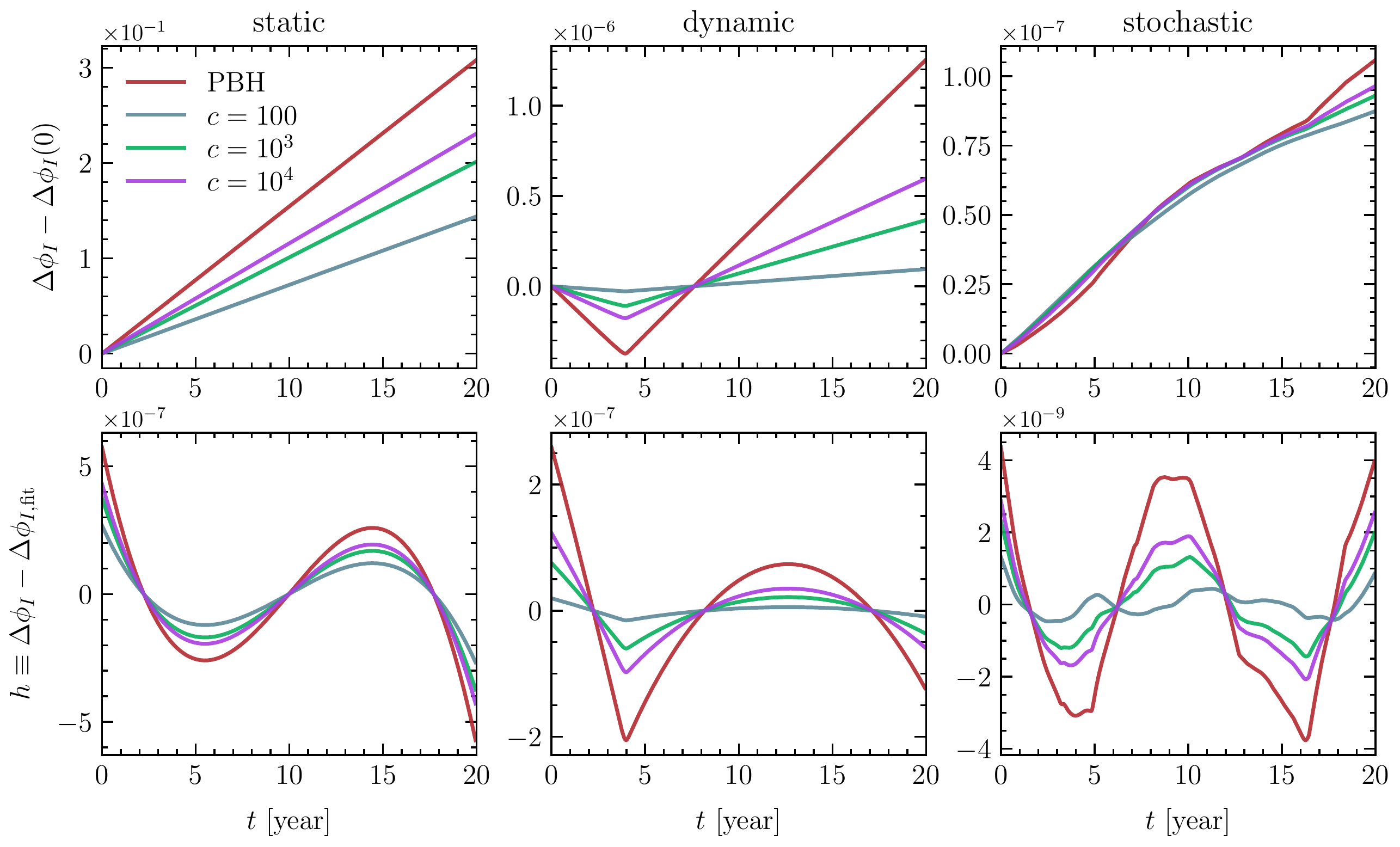}
    \caption{Sample signal shapes of the static (left column), dynamic (middle column) and stochastic (right column) signals from DM subhalos with $c = 100, 10^3, 10^4$ and the Primordial Black Hole limit $c \rightarrow \infty$. The top row shows the unsubtracted signal shapes, computed from Eqs.~\eqref{eq:delta_phi_form} and \eqref{eq:Delta_phi_general}, while the bottom row shows the subtracted signal shapes after the terms degenerate with the timing model have been removed. The first two columns show the phase shift $\delta \phi(t)$, Eq.~\eqref{eq:delta_phi_form}, from a single transiting subhalo and the last is the signal shape from multiple transiting subhalos, $\Delta \phi$, in Eq.~\eqref{eq:Delta_phi_general}. The signal shapes are taken from a realization of the Monte Carlo. The mass of the subhalos are $10^{-2}~M_\odot$, $10^{-11}~M_\odot$ and $10^{-14}~M_\odot$ for the static, dynamic and stochastic signals respectively. Only the closest subhalo is kept for the static and dynamic signals, while all subhalos are kept for the stochastic signal.}
    \label{fig:signal_shapes}
\end{figure}

This effect is now easily generalized to a PTA with $N_P$ pulsars, each of which is influenced by $N$ subhalos.  In this case the phase shift of the $I^{\rm th}$ pulsar is given by
\begin{align}
    \Delta \phi_I(t) = \sum_{i = 1}^N \delta \phi(t; M_i, c_i, \vec{r}_0^i, \vec{v}^i),
    \label{eq:Delta_phi_general}
\end{align}
where the sum runs over all of the DM subhalos affecting the $I^\text{th}$ pulsar. 

As discussed in detail in Ref.~\cite{Ramani:2020hdo}, the phase shift itself is not directly observable. This is because the shift induced by transiting DM subhalos can be partially degenerate with the phase shift induced by the `natural' evolution of the pulsar frequency, described by the timing model 
\begin{equation}
\phi(t) =\phi^0+\nu t+\frac{1}{2}\dot\nu t^2,
\label{eq:timing_model}
\end{equation}
where $\phi_0$, $\nu$, $\dot\nu$ are the phase offset, pulsar frequency, and its first time derivative. In general, if a DM signal is present, each pulsar measures a total residual phase shift, $s_I$, given by the sum of signal, $h_I$, and noise, $n_I$:
\begin{equation}\label{eq:signal}
s_I(t)\equiv h_I(t)+n_I(t)\,.
\end{equation}
The DM signal, $h_I$, is defined as the part of the phase shift, $\Delta\phi_I$, that is not absorbed by the pulsar timing model,
\begin{equation}\label{eq:subtraction}
h_I(t)\equiv\Delta\phi_I(t)-\Delta\phi_{I,\rm fit}(t),
\end{equation}
where 
\begin{equation}
 \Delta \phi_{I,\text{fit}} = \frac{1}{T} \sum_{n = 0}^2 f_n(t) \int_0^T f_n(t')  \Delta\phi_I(t') \, dt', 
    \label{eq:fit_signal}
\end{equation}
with $f_n = \sqrt{2n + 1}P_n(2t/T - 1)$, $P_n$ the n$^\text{th}$ Legendre polynomials, and $\phi(t)$ given by Eq.~\eqref{eq:timing_model}. Similarly, the post-fit residual noise, $n_I(t)$, is given by the intrinsic pulsar phase evolution that is not reabsorbed by the timing model.

Sample unsubtracted and subtracted signal shapes ({\em i.e.} before and after fitting for $\phi^0$, $\nu$, and $\dot \nu$) are shown in \Fig{fig:signal_shapes}. The left and middle columns show the signal shapes from a single subhalo. Depending on the timescale of the interaction, $\tau$, we classify the signals as either ``static" ($\tau \gg T$, left column) or ``dynamic" ($\tau \ll T$, right column). Both limits allow for simplifications of Eq.~\eqref{eq:delta_phi_form}. In the static limit $\delta \phi$ can be Taylor expanded in $t/\tau$. The total $\delta \phi$, in the top row, then looks nearly linear, while the measurable subtracted signal, the bottom row, is $\propto f_3(t)$, defined below Eq.~\eqref{eq:fit_signal}, due to the timing model fit of $\phi^0, \nu$ and $\dot{\nu}$. A dynamic signal, on the other hand, has a more characteristic signal shape. Expanding the term $\propto \vec{b} \cdot \vec{d}$ in Eq.~\eqref{eq:delta_phi_form} for small $\tau$ leads to $\delta \phi \propto |t - t_0|$, which explains the kink in the upper middle panel in \Fig{fig:signal_shapes}. This parameterization can be simplified further, since the terms $\propto t$ will be subtracted, to $\delta\phi(t)\sim (t-\overline{t})\Theta(t-\overline{t})$, and therefore the signal can be parameterized in terms of two variables: an amplitude and $\overline{t}$. Lastly, the stochastic signal in the right column is the sum of a large number of individual subhalos with $\tau \lesssim T$ and we see that the signal has no simple parameterization. 

\subsection{Constructing the SNR}

With the signal defined, we now discuss the signal significance relative to noise, or SNR. As done in Refs.~\cite{Dror:2019twh, Ramani:2020hdo} we will use a matched filter procedure \cite{Moore:2014lga, Smith:2019wny, Allen:1997ad}. As discussed previously, each pulsar in the PTA measures the residual phase shift, $s_I$, given in \eq{eq:signal} which is the sum of signal, $h_I$, and noise, $n_I$. Naively one would expect that the signal is only significant if $|h_I| > |n_I|$. However if we know the form of the signal, we can filter the noise and boost the significance. The optimal scenario for this filtering procedure is the one where the shape of the signal is known \textit{a priori}, or can be parameterized by a computationally searchable space of parameters. In this case the best test statistic is $\mathcal{T} \equiv \sum_I \int dt\,s_I(t) Q_I^{\rm opt}(t)$, where the signal of each pulsar is convolved with its own {\em optimal} filter, $\widetilde{Q}_I^\text{opt} = \widetilde{h}_I/\widetilde{N}_I$, where $\sim$ denotes the Fourier transform and $\widetilde{N}_I$ is the noise power spectrum of the $I^{\rm th}$ pulsar. We will assume the noise is white, so that we have
\begin{align}
    \widetilde{N}_I = \nu_I^2 t_\text{rms}^2 \Delta t \, ,
    \label{eq:pulsar_noise}
\end{align} 
where $t_\text{rms}$ is the root mean square residual timing noise, and $\Delta t$ is the time between measurements (the cadence). However to know, or find through a Markov Chain Monte Carlo, the optimal filter for all the pulsars in the array is not always possible.  We discuss alternatives for both the pulsar and Earth terms.  

\begin{figure}[ht]
    \centering
    \includegraphics[width=\textwidth]{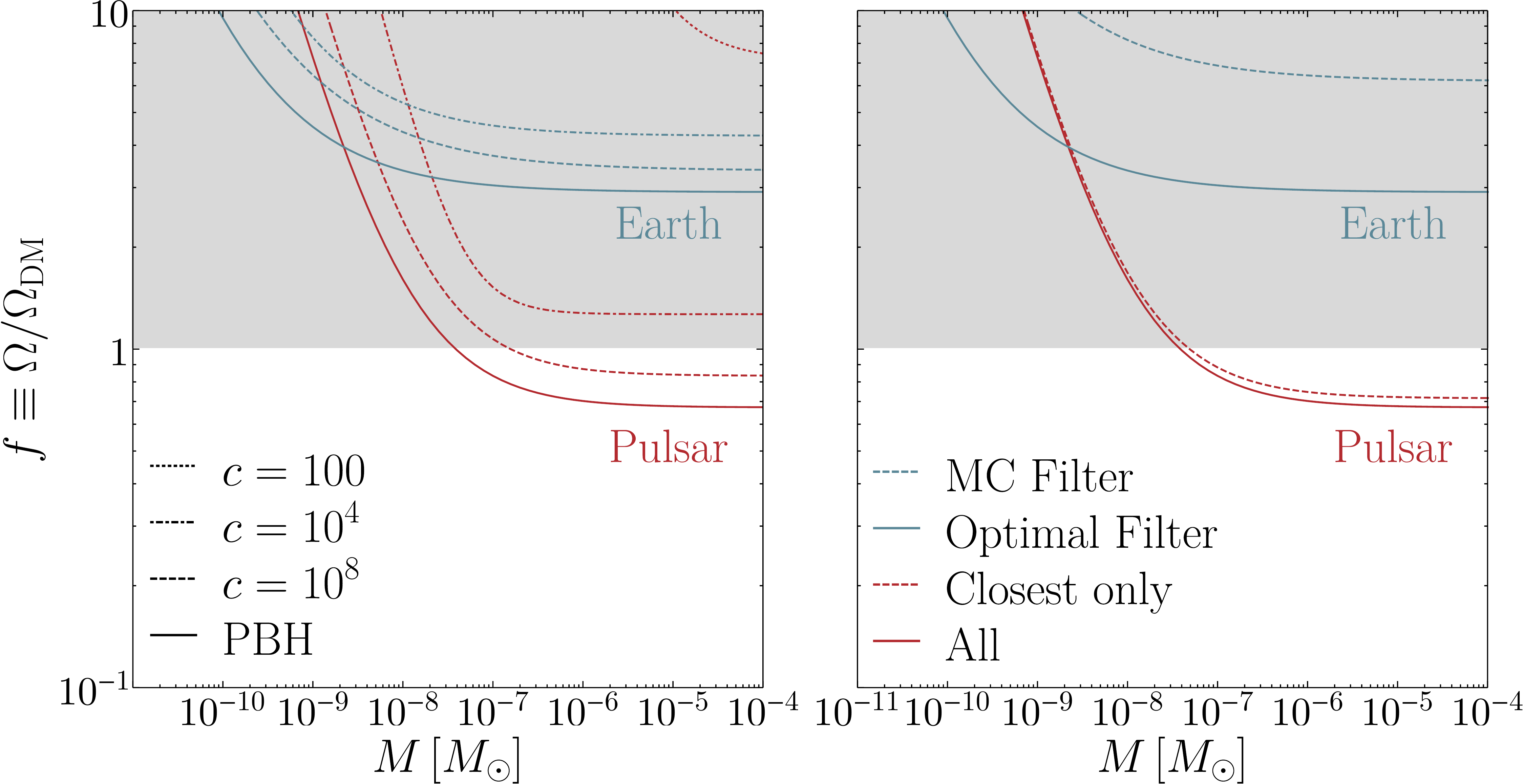}
    \caption{Limits from PTAs on the dark matter mass fraction $f$ = $\Omega/\Omega_{\mathrm{DM}}$ in subhalos of mass $M$. The left panel compares different subhalo concentration parameters, $c$ = $100$, $10^4$, $10^8$, and the PBH limit, $c\to \infty$, as well as the different search types, ``Earth" and ``Pulsar". The right panel shows how the constraints change when different assumptions are used when computing the SNR for the Earth and the Pulsar terms discussed in Sec.~\ref{sec:signal_deriv}. For the pulsar term we compare the constraints from keeping only the closest subhalo around each pulsar, labelled ``Closest only", to the constraint obtained from keeping all of the subhalo contributions, ``All". For the Earth term we compare the optimal filter to the one generated by the MC to understand the \textit{mis-filtering} effect. The PTA parameters used were $N_P=200$, $\Delta t=2\wk$, $t_{\mathrm{rms}}=50\ns$ and $T=20\yr$.}
    \label{fig:pbh_and_filter}
\end{figure}

We start by discussing the pulsar term. For a generic pulsar in the array, many transiting events contribute to the signal $s_I(t)$. This makes unfeasible to look for the optimal filter of a generic pulsar, given the large number of parameters needed to describe the signal shape. However, we expect the largest signal in the array to be dominated by a single halo transiting very close to the pulsar. Therefore, the largest signal is expected to have the simple parametrization discussed below \eq{eq:fit_signal}, which makes the search for its optimal filter feasible. We therefore define the {\em P}ulsar SNR, SNR$_P$, to be the largest across the array:
\begin{align}
    \text{SNR}_P^2 = \text{max}_I \left\{ \text{SNR}_I^2 \right\} = \text{max}_I \left\{ \frac{\left| \displaystyle\int df \, \widetilde{s}_I(f) \widetilde{Q}^*(f) \right|^2}{\displaystyle\int df \, \widetilde{N}_I(f) \widetilde{Q}^*(f) \widetilde{Q}(f)} \right\}  \, .
    \label{eq:s_snr_pulsar}
\end{align}
Concretely, to place a constraint on a model we would generate an array of filters with different values of the parameters in the signal model (e.g. an amplitude for the static signal, and an amplitude plus epoch, $\overline{t}$, for the dynamic signal), use these as the filter $Q$ in Eq.~\eqref{eq:s_snr_pulsar}, and then place constraints with the largest $\text{SNR}_P$.\footnote{This is not the only way the analysis could be done. Since the signal shapes are known the best fit signal parameters could be searched for with an Markov Chain Monte Carlo (MCMC), such as Enterprise \cite{Ellis2020}, and then compared to this distribution predicted by the MC presented here. We plan to explore this in a future work.} In practice, we assume we can select a filter which is arbitrarily close to $Q^\text{opt}$, and compute the expected maximum SNR from it,
\begin{align}
    \overline{\text{SNR}}_P^2 = \text{max}_I \left\{ \frac{1}{\widetilde{N}_I} \int dt \, h_I^2 \right\} \, .
    \label{eq:max_ex_snr_p}
\end{align}
The statistical significance of a given SNR, for the purposes of limit setting, is discussed in Appendix~\ref{app:snr_significance}. The $h_I$'s are generated with the MC which will be discussed in more detail below, as well as in Appendix~\ref{app:monte_carlo}. The resulting constraints for a monochromatic population of subhalos are labeled  ``Closest only" in the right panel of Fig.~\ref{fig:pbh_and_filter}, where only the signal from the closest subhalo is kept. To verify this approximation we also show the constraints obtained by taking $h_I$ to be the sum of the signals from all the subhalos around each pulsar, they are labelled ``All.'' in the right panel of Fig.~\ref{fig:pbh_and_filter}. 

The second signal, labelled ``Earth'' in the left panel of Fig.~\ref{fig:pbh_and_filter}, originates from a large flux of subhalos transiting near the Earth. Since the signal is generated by an acceleration of the Earth it will leave an imprint across the entire array with correlations between pulsars, similar to the correlations in the stochastic gravitational wave background \cite{Jenet:2014bea, Hellings:1983fr}. This signal is generated stochastically by many passing subhalos; as such, the optimal filter is not easily reconstructed. One can, however, parameterize the signal autocorrelator, $R_{IJ}(t, t') \equiv \langle h_I(t) h_J(t') \rangle$, where $\langle \rangle$ denotes an ensemble average. As in Ref.~\cite{Ramani:2020hdo}, we then define the SNR in terms of the signal correlator, $s_I(t) s_J(t')$,
\begin{align}
    \text{SNR}_E^2 = \frac{1}{2} \frac{\left| \sum_{I \neq J} \displaystyle\int df df'\, S_{IJ}(f, f') \widetilde{Q}^*_{IJ}(f, f') \right|^2}{\sum_{I \neq J} \displaystyle\int df df' \, \widetilde{N}_I(f) \widetilde{N}_J(f') \widetilde{Q}_{IJ}^*(f,f') \widetilde{Q}_{IJ}(f,f') },
    \label{eq:snr_earth_filter}
\end{align}
where $S_{IJ}(f, f') = \widetilde{s}_I(f) \widetilde{s}_J(f')$. If $\widetilde{s}_I(f)$ is known {\em a priori} then the optimal filter to use is $\widetilde{Q}_{IJ}^\text{opt} = \widetilde{s}_I(f) \widetilde{s}_J(f')/\left( \widetilde{N}_I \widetilde{N}_J \right)$. Since this is not the case, we use a filter based on the expected signal, $Q_{IJ} = \widetilde{R}_{IJ}/\left( \widetilde{N}_I \widetilde{N}_J \right)$, where $\widetilde{R}_{IJ} = \langle \widetilde{h}_I(f) h_J(f') \rangle$. Since the expected filter will not be optimal, this introduces a \textit{mis-filtering} effect. In the right panel of Fig.~\ref{fig:pbh_and_filter} we illustrate the effect that mis-filtering has on the constraints, lowering them by an $\mathcal{O}(1)$ number. We compute this constraint by using the MC to generate $S_{IJ}(f,f')$ in numerous realizations, taking the ensemble average $\widetilde{R}_{IJ}(f,f')$ over all the realizations, re-labelling it to $Q_{IJ}(f,f')$ and computing the SNR of the signal from new realizations using Eq.~\eqref{eq:snr_earth_filter}.  We will take the latter, more conservative, limits below. 

As discussed in detail in Appendix~\ref{app:monte_carlo}, to generate signals for monochromatic halo mass functions, as shown in Fig.~\ref{fig:pbh_and_filter}, the MC first generates the pulsar positions uniformly on a sphere. Then, for the Pulsar (Earth) term, around each pulsar (the Earth), subhalo initial positions $\vec{r}^0_i$ and velocities $\vec{v}_i$ are generated and used to compute the signal in Eq.~\eqref{eq:Delta_phi_general}. The initial positions are assumed to be uniformly distributed throughout space with density fixed to be $0.46$ GeV$/$cm$^3$, and the velocities drawn from a Maxwell-Boltzmann distribution with RMS velocity $v_0 = 325 \mbox{ km/s}$. The relevant SNR is then computed in 1000 realizations and constraints are placed setting the 10$^\text{th}$ percentile SNR equal to $4$ which, as detailed in Appendix~\ref{app:snr_significance}, correspond to a signal significance of $\sigma_{\rm significance} \sim 2$.

\section{From Primordial Perturbations to the Local Subhalo Population}
\label{sec:primordial_to_local}

As we have seen in the previous section, PTAs are a powerful tool to constrain the abundance of Milky Way (MW) DM subhalos. These sub-galactic structures are seeded by the primordial perturbations on scales much smaller than the ones tested by CMB observables, where DM models can leave unique fingerprints. In principle, once the power spectrum of primordial perturbations is known the abundance of these DM subhalos can be derived. In practice, however, this is an intricate problem where non-linearities, tidal effects, and baryonic feedback play an important role. For the $\Lambda$CDM model we can rely on numerical simulations which have been well-tested with semi-analytic fits. For other models of DM we need to employ motivated analytic estimates. The goal of this section is to illustrate the analytic prescription that we will adopt in the rest of this work to compute the subhalo mass function (sHMF) and concentration parameters.

We start by writing the dimensionless primordial power spectrum for DM density perturbations, $\Delta^2(k)$, as 
\begin{equation}
    \label{eq:primpow}
    \Delta^2(k)=\Delta_{\rm SI}^2(k)+\Delta_{\rm MD}^2(k)
\end{equation}
where $\Delta_{\rm MD}^2(k)$ is the model dependent contribution to the small scale power, and $\Delta_{\rm SI}^2(k)$ is the scale-invariant primordial power spectrum for adiabatic perturbations as extracted from CMB measurements (and extrapolated to smaller scales) \cite{Aghanim:2018eyx}:
\begin{equation}
    \Delta_{\rm SI}^2(k)=A_s\left(\frac{k}{k_0}\right)^{n_s-1},
\end{equation}
with $n_s=0.9665$, $A_s=2.101 \times 10^{-9}$, and pivot scale $k_0=5 \times 10^{-3} \Mpc^{-1}$. Following the standard assumption of linear evolution of the density perturbations, we relate this primordial power spectrum to its late-time value by means of transfer, $T(k)$, and growth, $D(z)$, functions:
\begin{equation}
    \label{eq:latepow}
    \Delta^2(k,z<z_{\rm eq})=T^2(k)D^2(z)\Delta_{\rm SI}^2(k)+T_{\rm MD}^2(k)D_{\rm MD}^2(z)\Delta_{\rm MD}^2(k),
\end{equation}
where $z_{\rm eq}\simeq 3.3\times 10^3$ is the redshift at matter-radiation equality, and we allow $\Delta_{\rm SI}$ and $\Delta_{\rm MD}$ to have different transfer and growth functions.

\begin{figure}[t!]
    \centering
    \textbf{Power spectrum at recombination}\par\medskip
    \includegraphics[width=\textwidth]{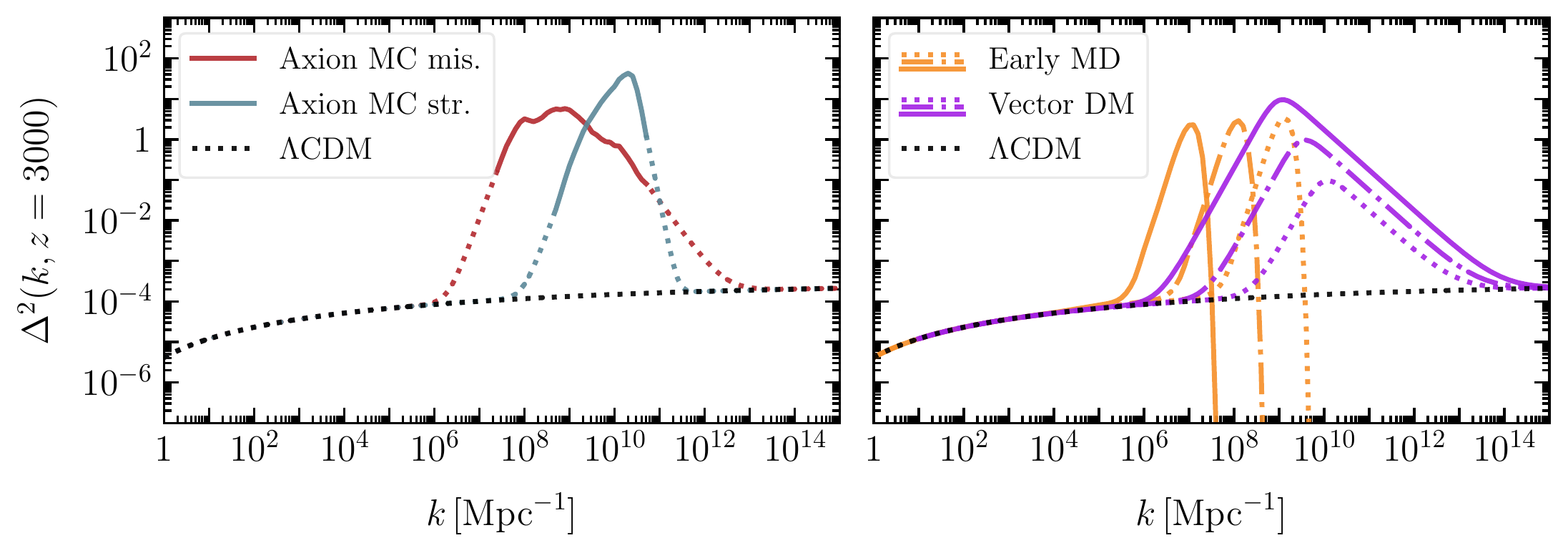}
    \caption{ \textbf{Left:}  dimensionless power spectrum at recombination for $\Lambda$CDM (black) compared to the post inflationary axion models (red for the misalignment dominated scenario, blue for the string dominated one). Dashed lines indicate regions where, due to lack of numerical or experimental results, the power spectra has been extrapolated. \textbf{Right:} similar to the left panel, except for the vector DM (purple), and early matter domination (orange) models replacing the post inflationary axion. For the early MD model we show the power spectrum obtained with three different choices of the reheating temperature: $T_{\rm RH}= 10\MeV$ (solid), $T_{\rm RH}= 0.1\GeV$ (dashdot), and $T_{\rm RH}= 1\GeV$ (dotted). For the vector DM model we fix the inflationary scale at $H_I=10^{14}\GeV$ and consider three DM masses: $m= 10^{-15}\GeV$ (solid), $m= 10^{-14}\GeV$ (dashdot), and $m= 10^{-13}\GeV$ (dotted).}
    \label{fig:models}
\end{figure}

The first category of models we consider modifies $\Lambda$CDM by increasing the amplitude of small-scale perturbations (\emph{i.e.} $\Delta^2_{\rm MD}(k)\ne 0$). The post-inflationary axion and vector DM models, discussed in Secs.~\ref{subsec:axion_MC} and \ref{subsec:vector:DM} respectively, belong to this category. The second category of models preserves the primordial power of $\Lambda$CDM (\emph{i.e.} $\Delta^2_{\rm MD}(k)=0$) but modifies the growth of small-scale perturbations in the early Universe, \emph{i.e.} an enhanced, non-standard $T(k)$. Models featuring an early stage of matter domination, discussed in Sec.~\ref{subsec:early_MD}, belong to this category. The net effect of both of these categories is similar: they enhance the power of matter density fluctuations on small scales, as shown in \Fig{fig:models}.  We will discuss these figures in more detail below, but simply note that the models under consideration feature a large enhancement of density fluctuations at scales well below what is currently measured with large scale structure surveys.

\subsection{Halo Mass Function}
The first quantity we derive from $\Delta^2(k,z)$ is the global Halo Mass Function (HMF), a function that gives the comoving number density of isolated halos in the Universe as a function of mass. Despite the non-linear nature of the problem, the analytic approach pioneered by Press \& Schechter \cite{Press:1973iz} has been proven to well-approximate numerical N-body simulations (at least for the standard $\Lambda$CDM). 

In the Press-Schechter (PS) formalism, overdense regions are expected to detach from the Hubble flow and gravitationally collapse when their average overdensity, $\delta \equiv \delta\rho/\rho$, becomes larger than a critical value, $\delta_c$ (which in a flat Universe is found to be $\delta_c\simeq1.686$).\footnote{For gravitational collapse during the radiation dominated epoch the critical overdensity becomes redshift dependent: $\delta_c(z)\simeq1.686 \left( \frac{3}{2} \frac{z}{z_{\rm eq}} \right)$ \cite{Kolb:1994fi, Ellis:2020gtq}.} Therefore, for Gaussian perturbations, the probability an overdense region of mass scale $M$ is collapsed by redshift $z$ is given by 
\begin{equation}
    \label{eq:PSprob}
    f(M,z)=\frac{2}{\sqrt{2\pi}\sigma(M,z)}\int_{\delta_c}^\infty d\delta\;e^{-\delta^2/2\sigma^2(M,z)}
\end{equation}
where $\sigma(M,z)$ is the variance of density perturbations smoothed over a sphere of size $R(M)=(3M/4\pi\rho_m)^{1/3}$:
\begin{equation}
    \sigma^2(M,z)=\int_0^\infty\frac{dk}{k}\,\Delta^2(k,z) W^2\left(k\,R(M)\right),
\end{equation}
where $W(x)$ is the window function of choice, and $\rho_m$ is the background matter density of the universe today. In the following, unless otherwise specified, we will use a top-hat window function, $W(x)=3 x^{-3}(\sin (x)-x \cos (x))$. 

The fact that a region of size $R(M)$ is collapsed does not prejudice against being part of a larger overdensity which has also collapsed. The HMF only counts collapsed objects which are not contained within larger halos; to obtain the comoving number density of these we differentiate Eq.~\eqref{eq:PSprob} with respect to $M$ and multiply by the average number density, $\rho_m/M$,
\begin{equation}
    \frac{dn(M,z)}{d\ln M} = \frac{\rho_m}{M}\frac{df(M,z)}{d\ln M}\,,
\end{equation}
with the mass fraction per logarithmic interval, $df/d\ln M$, is 
\begin{equation}
    \frac{df(M,z)}{d\ln M}=\sqrt{\frac{2}{\pi}}\nu(M,z)\exp\left(-\frac{\nu^2(M,z)}{2}\right)\frac{d\ln\sigma(M,z)}{d\ln M}\,,
\end{equation}
where we have defined $\nu(M,z)\equiv\delta_c/\sigma(M,z)$. 

The PS formalism has been validated against numerical simulations for $\Lambda$CDM cosmology \cite{Springel:2008cc, Fiacconi:2016cih}. But whether it is still a good approximation for models that, on small scales, differ vastly from $\Lambda$CDM has yet to be studied in detail. We plan to study this question in future work, but for the moment we limit our discussion to PS predictions and compare them to numerical results available for the case of axion miniclusters \cite{Eggemeier:2019khm}. The results of this comparison, shown in Appendix \ref{app:PS_comparison}, suggest that -- at least for axion miniclusters -- PS still provides a good estimate of the HMF. All the following results are based on this assumption. 

\subsection{Subhalo Mass Function}
As already mentioned, the target of PTA searches are pulsars within a radius of $\mathcal{O}(5\kpc)$ from our solar system. Because of this, PTAs can only detect signals generated by the local population of MW DM subhalos, and not from the cosmological population described by the HMF. Unfortunately, estimating the local subhalo density is a much more complicated problem than deriving the HMF. Indeed, in the galactic environment these subhalos are exposed to tidal forces that strip part of their mass, and which may ultimately lead to their complete disruption.

We divide the problem of deriving the subhalo mass function (sHMF), $d \tilde{n} / d \log{M}$, in two steps. First we derive its value at infall: $d\tilde{n}/d\log M_{\rm acc}$, where $M_{\rm acc}$ is the subhalo mass at the moment of its accretion into the MW and before tidal effects may reduce its value. Then, we analytically estimate the impact of tidal effects and derive the relation between the infall and final subhalo mass
\begin{equation}
    \label{eq:fbound1}
    M \equiv f_{\rm b}(M_{\rm acc})M_{\rm acc},
\end{equation}
where we have implicitly defined the bound fraction, $f_{\rm b}(M_{\rm acc})$, as the fraction of the infall subhalo mass that survives tidal effects. Finally, by using \eq{eq:fbound1}, we relate the infall sHMF to its final value,
\begin{equation}
    \frac{d\tilde n}{d\log M}=\int d\log M_{\rm acc}\frac{d\tilde n}{d\log M_{\rm acc}}\delta\left(1-\frac{f_{\rm b}(M_{\rm{acc}})M_{\rm acc}}{M}\right)\,,
\end{equation}
where here and in the following, we adopt a $\sim$ notation to distinguish between the local sHMF and the global HMF. 

As we will see, the density profile plays a key role in determining the impact of tidal effects. Subhalos are expected to have a radial density profile which is well described by the usual NFW profile given in Eq.~\eqref{eq:nfw_profile}. The characteristic density of the subhalo, Eq.~\eqref{eq:nfw_profile}, is set by the Universe's energy density at the time of collapse,
\begin{equation}
    \label{eq:rho_col}
    \rho_s=C_\rho \rho_c\left[\Omega_{\rm DM}(1+z_{\rm col})^3+\Omega_r(1+z_{\rm col})^4\right],
\end{equation}
where $C_\rho$ is a free parameter that encodes the growth of the halo density during the virialization processes, and $z_{\rm col}$ is the collapse redshift. We fix the value of $C_\rho$ by fitting \eq{eq:rho_col} against $N$-body simulations. For $\Lambda$CDM halos (for which $z_{\rm col} \simeq 10$) we use the results in \cite{Wang:2019ftp} and find $C_\rho \sim 600$, for axion miniclusters (for which $z_{\rm col}\gtrsim10^3$) we use the results of \cite{Eggemeier:2019khm} and find $C_\rho\sim9\times 10^4$. In the following we use a $z_{\rm col}$-dependent $C_\rho$ which smoothly interpolates between these two values.

\begin{figure}[t!]
\centering
\textbf{Tidal Effects}\par\medskip
\includegraphics[width=1\textwidth]{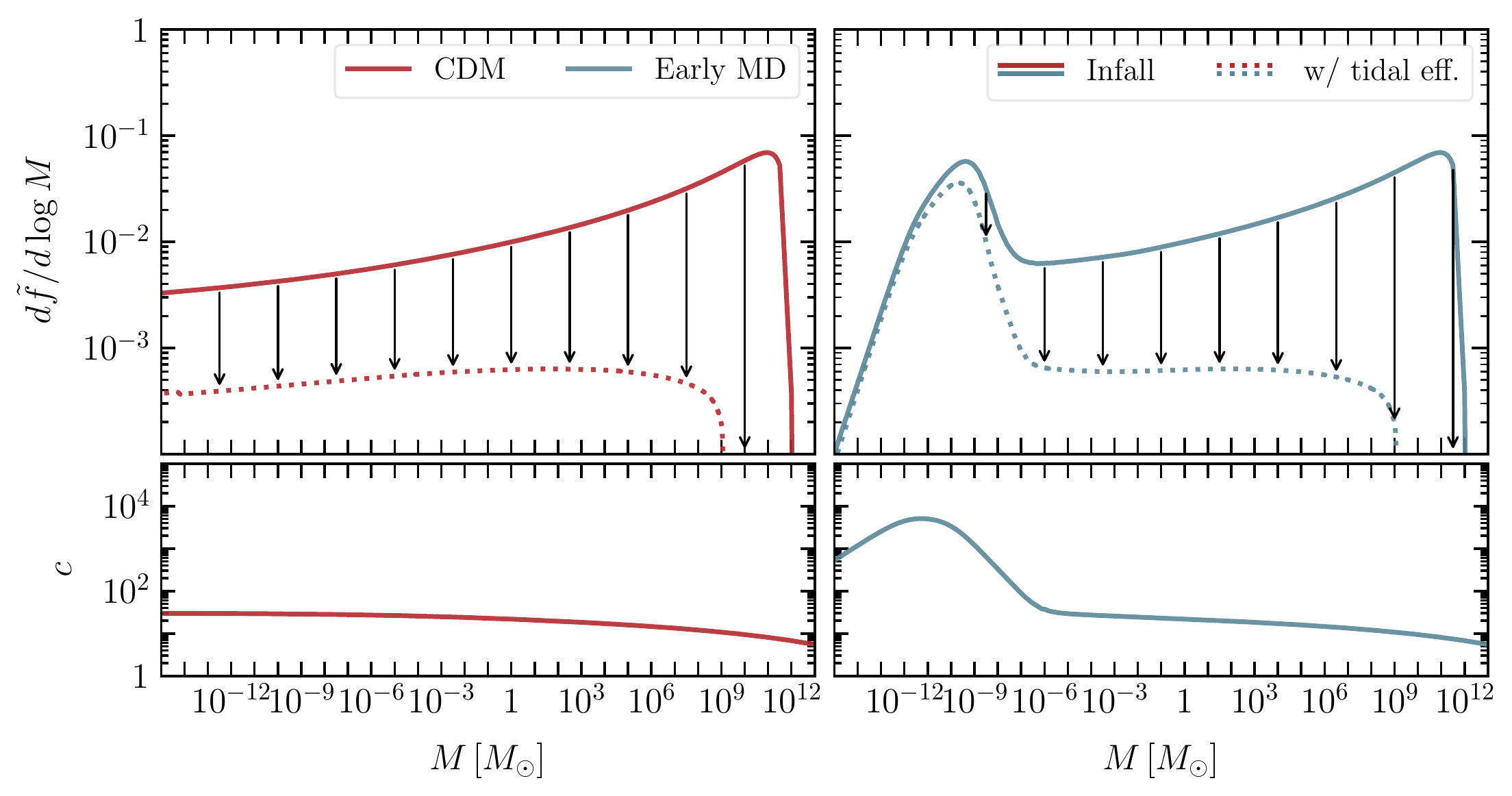}
\caption{Illustrative example of the impact of tidal effects on the subhalo mass fraction. {\bf Upper:} subhalo mass fraction before (solid) and after (dotted) tidal disruption for standard $\Lambda$CDM (red), and early MD (blue) with reheating temperature $T_{\rm RH}=10\MeV$ (for more details see Sec.~\ref{subsec:early_MD}). {\bf Lower:} typical concentration parameter as a function of the subhalo mass. Additional power on small scales leads to an early collapse and therefore a larger concentration parameter for light halos.}
\label{fig:df}
\end{figure}

Following Ref.~\cite{Navarro:1995iw}, we define $z_{\rm col}$ for a subhalo of final mass $M$ as the redshift at which half of its final mass is contained in progenitors more massive than $\epsilon M$. Using the extended Press-Schechter model \cite{Lacey1993}, this redshift can be estimated by
\begin{equation}
    \label{eq:zc}
{\rm erfc}\left[X(z_{\rm col})-X(0)\right]=\frac{1}{2}\,,
\end{equation}
where 
\begin{equation}
 X(z)=\frac{\delta_c}{\sqrt{2\left[\sigma^2(\epsilon M,z)-\sigma^2(M,z)\right]}},
\end{equation}
and the best fit to $\Lambda$CDM halos is found for $\epsilon=0.01$. In the following we will assume that the same value of $\epsilon$ provides a good estimate also for different DM models. The typical concentration parameters, as a function of mass, are shown in the lower panels of \Fig{fig:df}, where \eq{eq:rho_to_c} was used to relate $\rho_s$ to the concentration parameter. As expected, larger amplitude primordial perturbations lead to an earlier collapse and in turn larger concentration parameters.

\subsubsection{sHMF at infall}
In the PS formalism the probability that a halo of mass $M$ at redshift $z$ will be part of a larger halo of mass $M_0$ at redshift $z_0$ is given by \cite{Lacey1993, Giocoli:2007gf}
\begin{equation}
    \label{eq:dfsub}
\tilde f(M,z;M_0,z_0) = \frac{1}{\sqrt{2\pi}} \frac{\delta_c(z-z_0)}{\left(s-S\right)^{3/2}}\exp \left( -\frac{\delta_c^2(z-z_0)^2}{s-S} \right),
\end{equation}
where $s\equiv\sigma^2(M,0)$ and $S\equiv\sigma^2(M_0,0)$. Therefore, as a proxy for the sHMF at infall we can use
\begin{equation}
    \frac{d\tilde n}{d\log M_{\rm acc}}=\frac{\rho_{\rm DM}}{M_{\rm acc}}\frac{d\tilde f(M_{\rm acc},z_{\rm MW}; M_{200}^{\rm MW},0)}{d\log M_{\rm acc}},
\end{equation}
where $\rho_{\rm DM}$ is the local DM energy density, $M_{200}^{\rm MW}\approx1.5\times10^{12} M_\odot$ is the MW virial mass, and $z_{\rm MW}$ is the MW collapse time, derived from \eq{eq:zc}. 

\subsubsection{Tidal effects}
Subhalos in the MW can experience different kinds of tidal forces. Subhalos on nearly circular orbits are subjected to an almost constant tidal pull from the galactic halo. The effect of this gravitational pull is to strip away the halo mass contained beyond its \emph{tidal radius}, $r_t$, defined as the distance from the subhalo center at which the gravitational pull of the galaxy is stronger than the self-gravity of the subhalo \cite{vandenBosch:2017ynq}:
\begin{equation}
    \label{eq:rtidal}
r_t=r_{\oplus}\left[\frac{M_{\rm acc}(r<r_t)/M_{\rm MW}(r<r_\oplus)}{3-\frac{d\ln M_{\rm MW}}{d\ln R} |_{r_\oplus}}\right]
\end{equation}
where $r_\oplus$ is the radius of the Earth's circular orbit (assumed to be the distance between the solar system and the center of the galaxy, $8$ kpc), $M_{\rm MW}(r<r_\oplus)$ is the Milky Way mass enclosed within radius $r_\oplus$, and $M_\text{acc}(r<r_t)$ is the accreted mass within $r_t$. For an NFW profile, the enclosed mass is given by $M(r < r^*) = M\mathcal{F}(r^*/r_v,c)$, where $\mathcal{F}$ is defined in Eq.~\eqref{eq:phi_form}, and $M$ is the total mass. Assuming that the mass outside the tidal radius is instantaneously stripped away, the halo bound fraction surviving tidal stripping is\footnote{In reality, tidal stripping happens over several orbits. However, if compared with numerical simulations of $\Lambda$CDM halos, the instantaneous stripping assumption seems to be accurate up to $\mathcal{O}(1)$ factors \cite{vandenBosch:2017ynq}.}
\begin{equation}
    f_{\rm b}(M_{\rm acc})=\frac{M_{\rm acc}(r<r_t)}{M_{\rm acc}}\,.
    \label{eq:f_bound}
\end{equation}
The treatment of tidal effects performed in this analysis should be seen as a simple order of magnitude estimate whose main goal is to take into account the large tidal disruption suffered by the standard $\Lambda$CDM halos. To accurately estimate the smaller impact that tidal effects have on high concentration subhalos a more sophisticated analysis is needed. Specifically, additional processes like subhalo-subhalo encounters and interactions with MW stars can be the dominant disruption mechanism for high density subhalos, and need to be considered. A detailed study of these effects would require dedicated numerical simulation that we leave for future works. However, we do not expect these effects to change our results drastically, as also suggested by the recent results of Ref.~\cite{Kavanagh:2020gcy}.  

\section{Constraints on primordial power spectra}
\label{sec:constraints}

\begin{figure}[t!]
    \centering
    \includegraphics[width=0.8\textwidth]{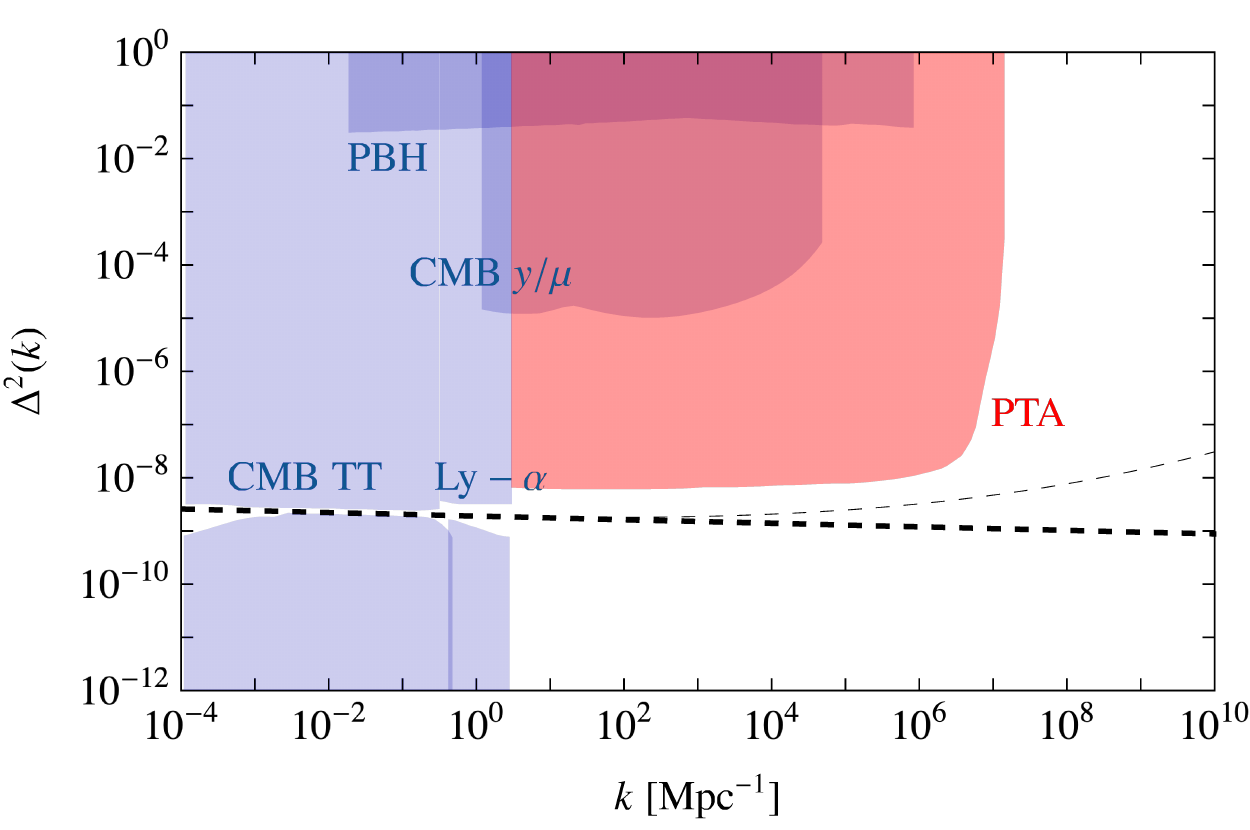}
    \caption{Limits on the dimensionless primordial power of curvature fluctuations. PTA limits (red) are compared to constraints from a combination of CMB \cite{Ade:2015lrj} and Ly-$\alpha$ \cite{Bird:2010mp} observables, together with limits on primordial black holes (PBH) (all in blue). The thick dashed line is the Planck best fit, assuming a constant spectral index, while the thin dashed line is the best fit obtained allowing the spectral index to have a $k$ dependence.}
    \label{fig:estimate}
\end{figure}
In this section we use the tools developed to derive the discovery potential of PTAs for four benchmark DM models. However before discussing model-specific results, we want to give a rough idea of the constraints that PTAs can place on the DM primordial power spectrum. Assuming this spectrum is locally scale invariant, it can be shown \cite{Bringmann:2011ut} that 
\begin{equation}
    \sigma^2(M,z) \sim \Delta^2(1/R(M),z)
\end{equation}
where, as before, $R$ and $M$ are related by $R(M)=(3M/4\pi\rho_m)^{1/3}$. This implies that there is an approximate one-to-one correspondence between the power of perturbations on scales $1/R(M)$ and the collapse probability, $f(M,z)$, given by \eqref{eq:PSprob}. Assuming that the population of subhalos is not drastically altered by their merger history, the fraction of the local DM energy density in subhalos of a given mass is $f(M,z)f_b(M)$, where $f_b(M)$ is the bound fraction, Eq.~\eqref{eq:f_bound}, which accounts for tidal effects that these subhalos experience once accreted into the MW halo. In general these subhalos will not be isolated objects but will form substructure of larger subhalos, meaning $f(M, z = 0)$ is certainly an overestimate of the mass fraction of subhalos of mass $M$. Bearing in mind these assumptions, we can say that PTA searches will be sensitive to a given amount of primordial power on scales $1/R(M)$ when
\begin{equation}
    f(M,z=0)f_b(M)>f_{\rm mono}(M,c)\, ,
\end{equation}
where the constraints for a monochromatic population of subhalos, $f_{\rm mono}(M,c)$, are shown in \Fig{fig:pbh_and_filter}.
Assuming the transfer function is the one predicted by $\Lambda$CDM, and $r_\oplus=8.2\kpc$, the constraints on the dimensionless primordial power are shown in \Fig{fig:estimate}. It is clear from this figure that PTA searches can strongly constrain models that have a small scale power spectrum that is enhanced compared to $\Lambda$CDM. However, standard $\Lambda$CDM subhalos, being almost completely disrupted by tidal effects (see \Fig{fig:df}), are a much more challenging target (as discussed in Sec. \ref{subsec:LCDM}).

\begin{figure}[t!]
\centering
\textbf{$\Lambda$CDM}\par\medskip
\includegraphics[width=\textwidth]{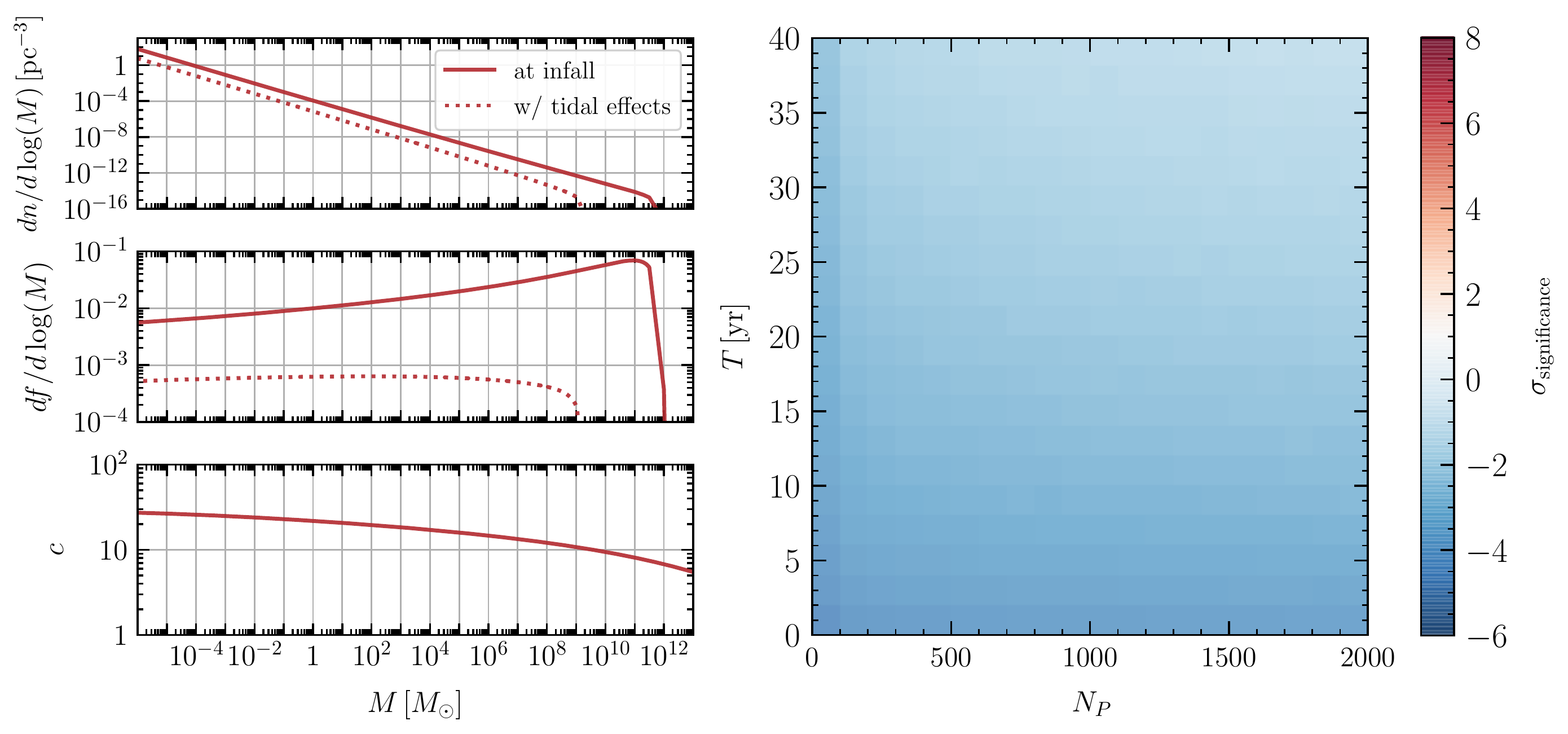}
\caption{\textbf{Left:} sHMF (top panel), and mass fraction per logarithmic interval (center panel) before (solid) and after (dashed) the inclusion of tidal effects. The lower panel shows the typical subhalo concentration parameter as a function of their mass. \textbf{Right:} Discovery significance for different values of the observation time and number of pulsars in the array. The residual timing noise and observation cadence have being fixed to $t_{\rm rms}=10\ns$ and $\Delta t=1\,{\rm week}$.}
\label{fig:CDM_results}
\end{figure}

\subsection{$\Lambda$CDM}
\label{subsec:LCDM}

The $\Lambda$CDM model is reproduced taking $\Delta^2_{\rm MD}(k)=0$ in \eq{eq:primpow}, and the standard growth and transfer functions in \eq{eq:latepow}. Specifically, we use the BBKS transfer function \cite{Bardeen:1985tr},
\begin{equation}
    T(x\equiv k/k_{\rm eq})=\frac{12 x^2}{5}\frac{\ln(1+0.171x)}{0.171x}\Big[1+0.284x+(1.18x)^2+(0.399x)^3+(0.490x)^4\Big]^{-1/4},\,
\end{equation}
and a linear growth function,
\begin{equation}
    \label{eq:lin_growth}
    D(z)=g(\Omega_m,\Omega_\Lambda)\frac{1+z_{\rm eq}}{1+z}.
\end{equation}
Here $a_{eq}\simeq3\times10^{-4}$ is the scale factor at matter radiation equality, $k_{\rm eq}\simeq0.01\Mpc^{-1}$ the largest scale that enters the horizon before equality, and we use the approximated growth factor given in \cite{Carroll:1991mt}:
\begin{equation}
    g(\Omega_m,\Omega_\Lambda)\approx\frac{5\,\Omega_m}{2\left[\Omega_m^{4/7}-\Omega_\Lambda+(1-\Omega_m/2)(1+\Omega_\Lambda/70)\right]},
\end{equation}
where $\Omega_m$ and $\Omega_\Lambda$ are the matter and vacuum energy densities, normalized to the critical density, $\rho_c$, respectively. 

The sHMF for standard CDM halos has been computed in the mass range $(10^{-5}-10^{12}) M_\odot$ \cite{Wang:2019ftp} but little is known for lighter subhalos. Because of this, we compute the sHMF at infall by using Eq.~\eqref{eq:dfsub}. As expected, the result is compatible with the numerical simulations, and the sHMF at infall is well fitted by
\begin{equation}
    \label{eq:LCDM_HMF1}
\frac{d\tilde n}{d\log M_{\rm acc}}=\mathcal{N}(M_{\rm max},M_{\rm min}) \left(\frac{M_{\rm acc}}{M_0}\right)^\alpha\Theta(M_{\rm max}-M_{\rm acc})\Theta(M_{\rm acc}-M_{\rm min}),
\end{equation}
with a slope of $\alpha=-0.95$, truncated at the DM free streaming scale, $M_{\rm min}$, and a host mass $M_{\rm max}\approx10^{12}M_\odot $. The normalization constant
\begin{equation}
\mathcal{N}(M_{\rm max}, M_{\rm min})=\frac{\rho_{\rm DM}}{1+\alpha}\left(M_{\rm max}^{1+\alpha}-M_{\rm min}^{1+\alpha}\right)\,
\end{equation}
is derived by requiring that, before tidal disruption takes place, DM subhalos account for all the local DM energy density. The final sHMF, including tidal effects, is derived following the prescription described in the previous section; the results of this procedure are shown in the left panels of \Fig{fig:CDM_results}.

We then use the MC to simulate the PTA signal that this population of DM subhalos would induce in a PTA. The statistical significance of this signal is shown in the right panel of \Fig{fig:CDM_results} for different values of the observation time and number of pulsars in the array, while the timing residual noise and cadence have been fixed to $t_{\rm rms}=10\ns$ and $\Delta t=1$ week. Despite the optimistic PTA parameters, the strong impact of tidal effects, which almost completely erase the local population of $\Lambda$CDM subhalos, makes the model impossible to test in present and future PTA experiments. 

\begin{figure}[t!]
\begin{center}
\textbf{Axion Miniclusters}\par\medskip
\includegraphics[width=\textwidth]{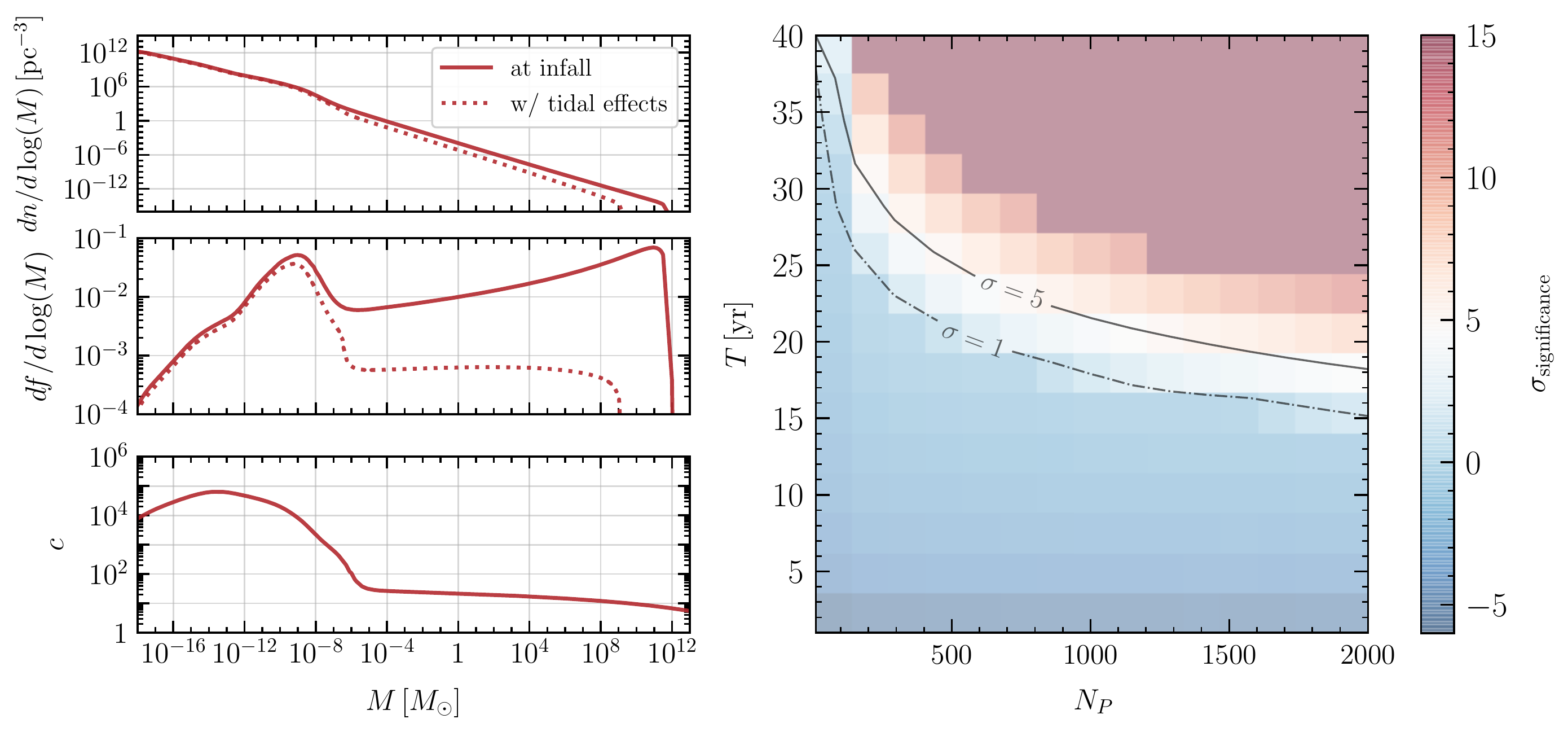}
\caption{\textbf{Left:} sHMF (top panel), and mass fraction per logaritmich interval (center panel) before (solid) and after (dashed) the inclusion of tidal effects. The lower panel shows the typical subhalo concentration parameter as a function of their mass. \textbf{Right:} Discovery significance for different values of the observation time and number of pulsars in the array. The residual timing noise and observation cadence have being fixed to $t_{\rm rms}=10\ns$ and $\Delta t=1\,{\rm week}$. The dashed (solid) line shows the $1\sigma$ ($5\sigma$) significance contour.}
\label{fig:axion_results}
\end{center}
\end{figure}

\subsection{Axions with PQ Symmetry Breaking After Inflation}
\label{subsec:axion_MC}
If the PQ symmetry breaks after inflation ($f_a \lesssim H_I$), the Universe is populated by casually disconnected patches each containing different values of the axion field. These patches are separated by a network of strings and domain walls which evolve through axion radiation until the QCD phase transition. At this point the PQ symmetry is explicitly broken, the network of topological defects decays, and the axion starts to oscillate around its minimum. From this point on the axion behaves as CDM and perturbations in the density field evolve as those of a collisionless fluid. 

The evolution of the PQ field down to the QCD phase transition has been studied numerically in \cite{Vaquero:2018tib, Buschmann:2019icd}. These simulations need to resolve string cores and contain a few Hubble patches at the same time. These two scales are respectively fixed by the mass of the radial mode, $m_r \approx f_a$, and $1/H \approx M_{Pl}/T^2$. At the QCD phase transition the scale separation is $\log\small(M_{Pl} f_a/\Lambda_{\rm QCD}^2\small) \approx 70$, which makes numerical progress to evolve from the PQ to QCD phase transition nearly impossible. Luckily strings are expected to enter a scaling regime, during which their length per Hubble patch remains constant, soon after the PQ phase transition. Therefore simulations can be stopped for reasonable values of $\log(m_r/H)$ and extrapolated to the QCD era. However, the authors of \cite{Gorghetto:2020qws} recently pointed out that large logarithmic corrections to this scaling regime are present, the extrapolation of which suggests that axions from strings may dominate the axion relic density contrary to what is assumed in \cite{Vaquero:2018tib, Buschmann:2019icd}. This would not only change the predicted axion mass but also the spectrum of its primordial perturbations. In the following we will set constraints assuming that the axion relic density is dominated by the misalignment contribution, with the possibility of updating the results in case the conclusion of reference \cite{Gorghetto:2020qws} is confirmed. 

Given this assumption we use the primordial spectrum for the axion field derived in \cite{Vaquero:2018tib}. Since isocurvature perturbations are expected to experience a very small logarithmic growth during radiation domination, and linearly evolve afterwards, in \eq{eq:latepow} we take
\begin{equation} 
    T_{\rm MD}(k)\approx 1\qquad{\rm and }\qquad D_{\rm MD}(z)\simeq\left(1+\frac{3}{2}\frac{1+z_{\rm eq}}{1+z}\right)\,,
\end{equation}
while for the transfer and growth function of the scale invariant part of the spectrum ($T(k)$ and $D(z)$) we use the same conventions described in the previous subsection. The properties of the resulting local population of DM subhalos are shown in the left panels of \Fig{fig:axion_results}. 

As before we use the MC to simulate the local population of DM subhalos and derive the induced PTA signal with the methods discussed in Sec.~\ref{sec:signal_deriv}, and the results are shown in the right panel of \Fig{fig:axion_results}. For SKA parameters the detection of axion MC appears to be challenging, but more futuristic PTA experiments should be able to test this model.  

Similarly to the axion case, models featuring a cosmological phase transition in the early Universe can boost the DM power spectrum on small scales \cite{Zurek2006} and be tested by PTAs. Interestingly, if the phase transition happens at low enough temperatures ($T\lesssim 1\GeV$), these models are also expected to generate a background of gravitational waves which could be searched for in PTAs (see for example \cite{Ratzinger2020}). We leave to future works a detailed study of these scenarios. 

\begin{figure}[t!]
\centering
\textbf{Early Matter Domination}\par\medskip
\includegraphics[width=\textwidth]{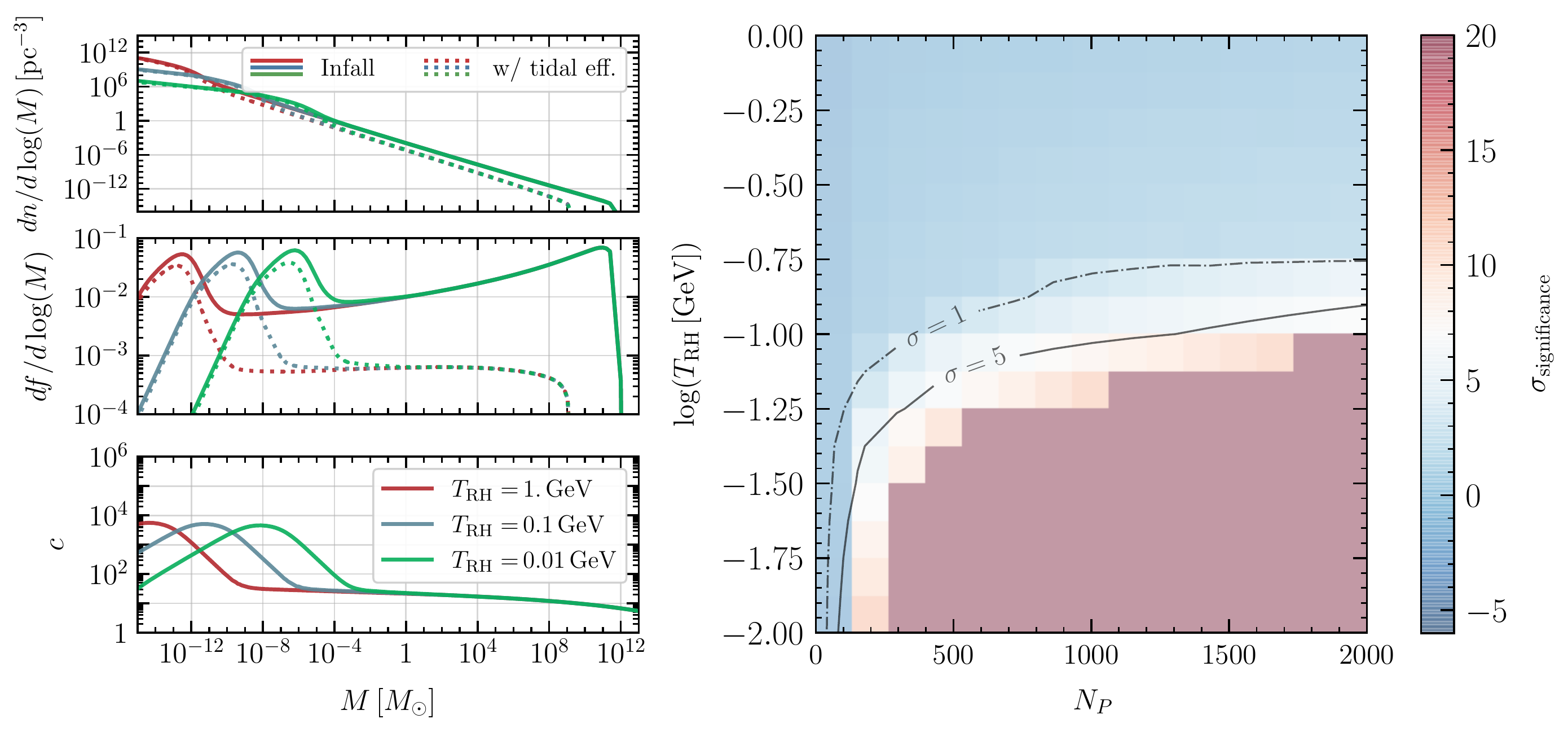}
\caption{\textbf{Left:} sHMF (top panel), and mass fraction per logarithmic mass interval (center panel) before (solid) and after (dashed) the inclusion of tidal effects. The lower panel shows the typical subhalo concentration parameter as a function of subhalo mass. All curves are shown for three values of the reheating temperature, $T_\text{RH} = 10 \MeV, 100 \MeV, 1 \GeV$. \textbf{Right:} Discovery significance for different values of the reheating temperature and number of pulsars in the PTA. The residual timing noise, observation cadence, and observation time are $10\ns$, $1\,{\rm week}$, and $30 \,{\rm yr}$, respectively. The dashed (solid) line shows the $1\sigma$ ($5\sigma$) significance contour.}
\label{fig:md_results}
\end{figure}

\subsection{Early matter domination}\label{subsec:early_MD}

Models with an early stage of matter domination (MD) are another class of models potentially testable by PTAs. Here, after the end of inflation, the Universe's energy density is dominated by a non-relativistic spectator field. Sub-horizon perturbations of this field then grow linearly with the scale factor during this early stage of MD. When the spectator field decays and reheats the Universe, these density fluctuations are inherited by the DM density field. This primordial growth of matter perturbations can be parametrized in terms of a modified matter transfer function, $T_\text{MD}$. Following ~\cite{Erickcek:2011us}, we include a period of early MD by modifying the transfer function,
\begin{equation}\label{eq:md_transfer}
    T_\text{MD}(k)=R(k)T(k)\exp\left(-\frac{k^2}{2k_{\rm fs}^2}\right)\,,
\end{equation}
where $T(k)$ is the standard BBKS transfer function, $R(k)$ encodes the modifications induced by the early matter domination era, and the exponential is given by the free streaming scale, induced by the finite velocity of the DM when it is produced from the decay of the spectator field. Large scales are not affected by the early matter domination so $R(k<0.05\;k_{\rm RH})=1$, where $k_{\rm RH}$ is the wave number of the mode that enters the horizon at reheating,
\begin{equation}
    \frac{k_{\rm RH}}{k_{\rm eq}}=1.72\times10^{11}\Bigg(\frac{T_{\rm RH}}{100\GeV}\Bigg)\Bigg(\frac{g_*}{100}\Bigg)^{1/6}\,,
\end{equation}
with $k_{\rm eq} \simeq 0.01\Mpc^{-1}$ the largest scale that enters the horizon before matter-radiation equality, $T_{\rm RH}$ the temperature at which the spectator field decays (\emph{i.e.} the reheating temperature), and $g_*$ the number of relativistic species at the time of reheating. 
On small scales, $k> 0.05\;k_{\rm RH}$,  early matter domination enhances the growth of perturbations such that \cite{Erickcek:2011us},
\begin{equation}
    R(k > 0.05\;k_{\rm RH})=\frac{A(k/k_{\rm RH})\ln\left[\left(\frac{4}{e^3}\right)^{f_2/f_1}\frac{B(k/k_{\rm RH}) a_{\rm eq}}{a_{\rm hor}(k)}\right]}{9.11\ln\left[\left(\frac{4}{e^3}\right)^{f_2/f_1}0.594\frac{\sqrt2 k}{k_{\rm eq}}\right]},
\end{equation}
with $a_{\rm hor}(k)$ the scale factor at reheating and horizon crossing, and $f_1$ and $f_2$ related to the baryon fraction $f_{\rm b}\equiv\rho_{\rm b}/(\rho_{\rm b}+\rho_{\rm dm})$ by
\begin{equation}
f_1=1-0.568 f_{\rm b}+0.094 f_{\rm b}^2\qquad
f_2=1-1.156 f_{\rm b}+0.149 f_{\rm b}^2-0.074f_{\rm b}^3\,.
\end{equation}
On small scales, the functions $A(x)$ and $B(x)$ are fit according to
\begin{equation}
\begin{split}
A(x)&= \exp \left[\frac{0.609}{\left(1+2.15 (\log x-1.52)^2\right)^{1.38}}\right]\left[9.11\,\mathcal{S}(5.02\, -x)+\frac{3}{5} x^2 \mathcal{S}(x-5.02)\right]\\
B(x)&=\exp\left[\log (0.594) \mathcal{S}(5.02\, -x)+\mathcal{S}(x-0.52) \log \left(\frac{e}{x^2}\right)\right]\,,
\end{split}
\end{equation}
where $\mathcal{S}(y)=[\tanh(y/2)+1]/2$.
Finally, ignoring DM interactions, the free-streaming scale, $k_{\rm fs}$, appearing in \eq{eq:md_transfer} is approximately given by
\begin{equation}
    \frac{k_{\rm RH}}{k_{\rm fs}}\approx  \frac{\langle v_{\rm RH}\rangle}{0.06}
\end{equation}
where $v_{\rm RH}$ is the DM velocity at reheating. In deriving our results we have assumed that $\langle v_{\rm RH}\rangle=10^{-3}$. For smaller values of $\langle v_\text{RH} \rangle$ the results shown in Fig.~\ref{fig:md_results} remain almost unchanged. However if the DM is produced with $\langle v_\text{RH} \rangle \gtrsim 0.06$ then $k_\text{RH} \gtrsim k_\text{fs}$ and free streaming erases the small scale perturbations of the scalar field entirely. In this case PTAs will not be able to set constraints. 

The properties of the subhalo population resulting from this modified transfer function are shown in the left panels of \Fig{fig:md_results}. While the significance of the predicted signal is reported on the right panel of the same figure. We see that, assuming optimistic PTA parameters, the model can be tested for $T_{\rm RH}\lesssim1\GeV$. For higher reheating temperatures typical subhalo masses become too small to induce a large enough signal and sensitivity is lost. 


\begin{figure}[t!]
\centering
\textbf{Vector Dark Matter}\par\medskip
\includegraphics[width=\textwidth]{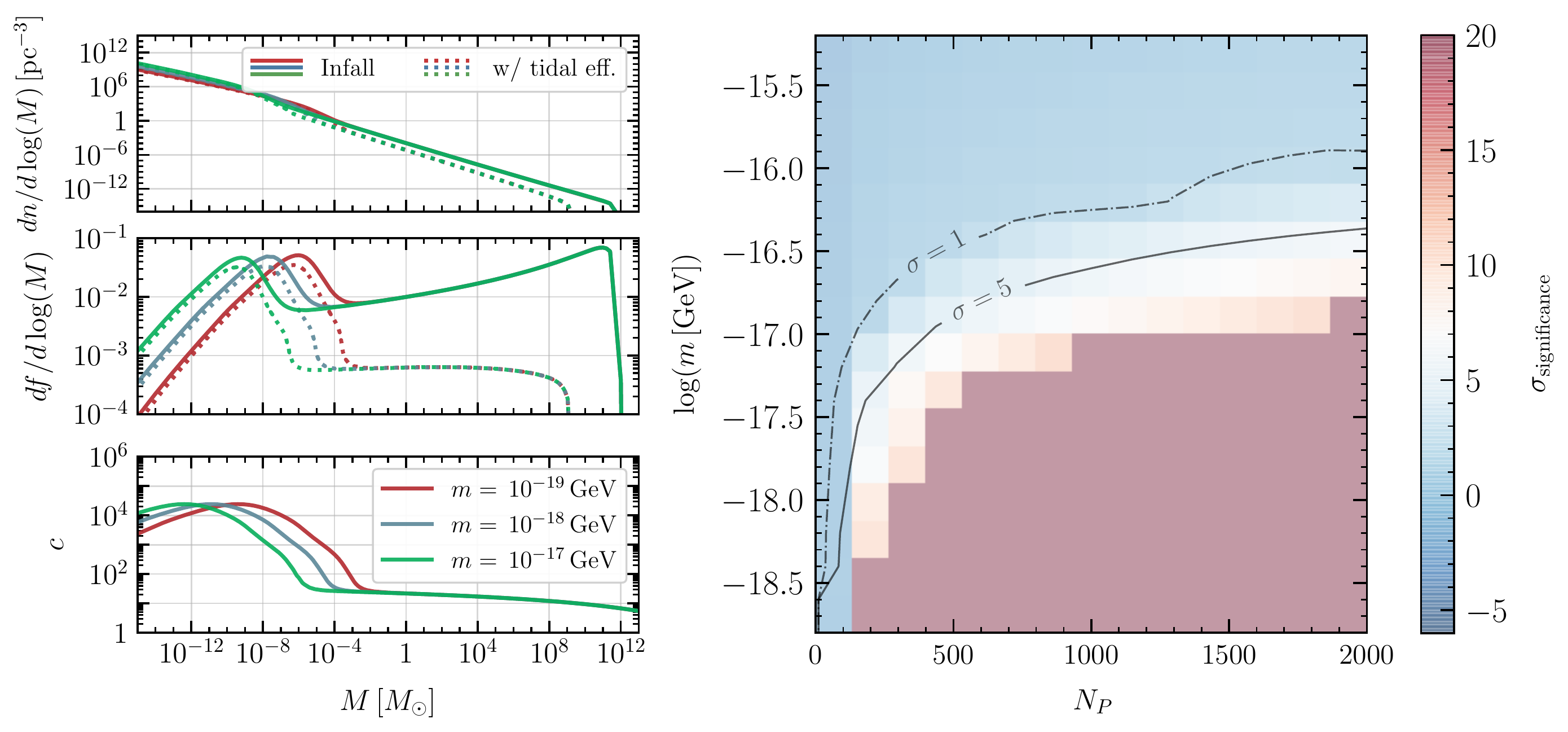}
\caption{\textbf{Left:} sHMF (top panel), and mass fraction per logarithmic mass interval (center panel) before (solid) and after (dashed) the inclusion of tidal effects. The lower panel shows the typical subhalo concentration parameter as a function of their mass. All curves are shown for three values of the DM mass, $m = 10^{-17}, 10^{-18}, 10^{-19} \GeV$. \textbf{Right:} Discovery significance for different values of the DM mass and number of pulsars in the array. The inflationary scale, $H_I$, has been fixed in order to reproduce the DM relic density according to \eq{eq:relic_vect}. The residual timing noise, observation cadence, and observation time have being fixed to $t_{\rm rms}=10\ns$, $\Delta t=1\,{\rm week}$, and $T= 30 \,{\rm yr}$. The dashed (solid) line shows the $1\sigma$ ($5\sigma$) significance contour.}
\label{fig:vdm_results}
\end{figure}

\subsection{Vector DM}\label{subsec:vector:DM}
As shown in \cite{Graham:2015rva}, massive vector bosons can be produced by quantum fluctuations during inflation. The relic abundance of particles produced in this way is given by
\begin{equation}
    \label{eq:relic_vect}
    \Omega\simeq\Omega_{\rm DM} \sqrt{\frac{m}{6\times 10^{-6}\eV}}\left(\frac{H_ I}{10^{14}\GeV}\right)^2,
\end{equation}
where $\Omega_{\rm DM}\simeq0.265$ is the observed DM relic density, $m$ is the mass of the boson, and $H_I$ is the Hubble rate during inflation. In addition to the adiabatic and scale invariant fluctuations inherited from perturbations in the inflaton field, the inflationary production generates isocurvature fluctuations on small scales. Specifically, the primordial power spectrum of the field amplitude is \cite{Graham:2015rva}
\begin{equation}
    \label{eq:vdm_D0}
    \Delta^2_{A}(k)=\left(\frac{k H_I}{2\pi m}\right)^2\,.
\end{equation}
Notice that this is not the power spectrum for DM density perturbations; we will discuss shortly how the two are related. At late times the small-scale power spectrum for the field amplitude has the usual relation
\begin{equation}
 \Delta^2_A(k,z<z_{\rm eq.})= T_A^2(k) D^2(z) \Delta^2_{A}(k),
\end{equation}
where $D(z)$ is the standard growth function, while the transfer function is 
\begin{equation}
\tilde T_A(k)=\sqrt{\frac{k_*}{m\,a_{\rm eq}}}\times
\left\{\begin{array}{ll}
\quad1&{\rm for}\; k<k_*\\ \\
\left(\dfrac{k_*}{k}\right)^{3/2}&{\rm for}\;k>k_*
\end{array}\right.\,,
\end{equation}
and $k_*=a_{\rm eq}\sqrt{H_{\rm eq} m}$, with $a_{\rm eq}=3\times 10^{-4}$ and $H_{\rm eq}\simeq10^{-29}\eV$ the value of the scale factor and Hubble rate at equality. Finally, the small-scale power spectrum for density perturbations is given by \cite{Graham:2015rva}:
\begin{equation}
    \Delta^2_{\rm MD}(k,z)=\frac{k^2}{4\langle A^2\rangle^2}\int_{|q-k|<p<q+k}\frac{(k^2-q^2-p^2)^2}{q^4p^4}\Delta^2_A(p,z)\Delta^2_A(q,z)\,dpdq
\end{equation}
where $\langle A^2\rangle=\int d\ln k\,\Delta_A^2(k,z)$.

In \Fig{fig:vdm_results} we show the properties of the subhalo population predicted by this model (left panel), and the significance of the signal that it would produce (right panel). We see that future PTA experiments will have a good sensitivity for DM masses below $10^{-16}\GeV$.

\section{Conclusions}\label{sec:conclusions}

We have studied the detectability  in Pulsar Timing Arrays of a variety of well motivated DM models of substructure including: standard $\Lambda$CDM, axion models where the PQ symmetry breaks after inflation, early matter domination, and vector DM. Given the low concentration, $\Lambda$CDM subhalos are particularly susceptible to tidal effects which drastically reduces their local abundance. As a result, we found that $\Lambda$CDM will not be testable by present or (near-)future PTAs. The other models, which feature subhalos with much larger concentration parameters, and hence much lower tidal disruption, are better candidates for detection. Specifically, we found that models featuring an early stage of matter domination ending at temperatures lower than $1\GeV$ will be testable by PTAs with SKA-like capabilities. Similarly, vector DM candidates produced during inflation with a mass smaller than $10^{-16}\GeV$ are within the reach of PTAs with SKA-like parameters. Finally,  axions whose PQ symmetry breaks after inflation (if the production is dominated by misalignment) are out of reach for an SKA-like experiment, but could be probe by a slightly more optimistic set of experimental parameters.

To generate the signals we have developed a Python Monte Carlo tool that, given the subhalo mass function, DM velocity distribution, and concentration parameters, generates a population of subhalos and computes the acceleration of the Earth, or pulsar, induced, and the resulting shift on the phase of the pulse time-of-arrival. We make the code publicly available on GitHub \githublink. 

In future works we plan to improve the present analysis in two ways. First, by performing dedicated N-body simulations to more precisely describe the impact of tidal effects on high-density subhalos. Second, by performing a more realistic treatment of the background noise through the NANOGrav software Enterprise \cite{Ellis2020}. 

\subsection*{Acknowledgments}

We thank Phil Hopkins, Stephen Taylor, and Huangyu Xiao for helpful discussions. This work is supported by the Quantum Information Science Enabled Discovery (QuantISED) for High Energy Physics (KA2401032).

\appendix

\section{Signal Generation Monte Carlo}
\label{app:monte_carlo}

Our goal is to compute the impact of a population of subhalos on the timing residuals measured in a PTA. The signal from an individual subhalo, given in Eq.~\eqref{eq:delta_phi_form}, is a function of a few random variables, specifically: the subhalo mass, $M$, initial position, $\vec{r}^0$, and velocity, $\vec{v}$. To generate the full signal all of these variables must be generated for each subhalo. We assume that these probability distributions are independent and identically distributed (iid). With these assumptions the population can be generated by drawing random variables from three probability density functions (pdfs):
\begin{itemize}
    \item $f_{\vec{r}}(\vec{r})$: the subhalo spatial distribution;
    \item $f_\vec{v}(\vec{v})$: the subhalo velocity distribution;
    \item $f_M(M)$: the subhalo mass distribution (related to the subhalo mass function, $d \tilde{n}/d\log{M}$ as described below),
\end{itemize}
all of which are normalized to 1: $\int dX f_X(X) = 1$. Since PTA searches are only sensitive to DM subhalos in the neighborhood of the solar system, we expect the position distribution to be uniform. Therefore we take $f_\vec{r} = 1/V$, where $V$ is the simulation volume. Practically this volume is limited by the total number of subhalos that can be kept in the simulation. The velocity distribution, $f_\vec{v}(\vec{v})$, is taken to be a Maxwell-Boltzmann distribution with $v_0 = 325$ km$/$s, $v_\text{esc} = 600$ km$/$s, and the angular dependence assumed to be isotropic. The larger value of $v_0$ relative to the standard choice of $\sim 200$ km$/$s is chosen to account for the Earth or pulsar velocity relative to the Galactic rest frame.\footnote{This also introduces an anisotropy in the velocity distribution which creates spurious finite volume effects in the simulation. Since we do not expect the anisotropy to change our results significantly we ignore this effect.} Lastly, the mass distribution can be written in terms of the subhalo mass function as
\begin{align}
    f_M(M) = \frac{1}{\overline{n}M} \frac{d \tilde{n}}{d\log{M}} \, ,
    \label{eq:app_dist_hmf}
\end{align}
where $\overline{n} \equiv \int dM d\tilde{n}/dM$. The concentration parameters are then taken from the concentration-mass relationship, $c(M)$, as discussed in Sec.~\ref{sec:constraints}. We also generate the pulsar directions, $\hat{\vec{d}}$ in Eq.~\eqref{eq:delta_phi_form}, from a uniform distribution on the sphere. 

Given these pdfs, the Monte Carlo generates a large number (taken to be $1000$ in our results) of different universe realizations and in each computes the total phase shift, Eq.~\eqref{eq:Delta_phi_general}, due to subhalos surrounding the Earth or pulsar. The total phase shift is computed on a uniform grid of time points with spacing equal to the cadence, $\Delta t$, and total number of points equal to $T/\Delta t$. The residual signal in each pulsar, $h_I$, is then computed by finding the best fit  to the parameters $\phi^0, \nu$, and $\dot{\nu}$ in the timing model of \eq{eq:timing_model} and subtracting the best fit timing model from the total phase shift. These $h_I$'s are then used to compute the SNRs discussed in Sec.~\ref{sec:signal_deriv} for each realization. To draw constraints we take the $10^\text{th}$ percentile SNR across universes. The statistical significance of a given SNR is discussed below in App.~\ref{app:snr_significance}. 

\begin{figure}[t]
    \centering
    \includegraphics[width=\textwidth]{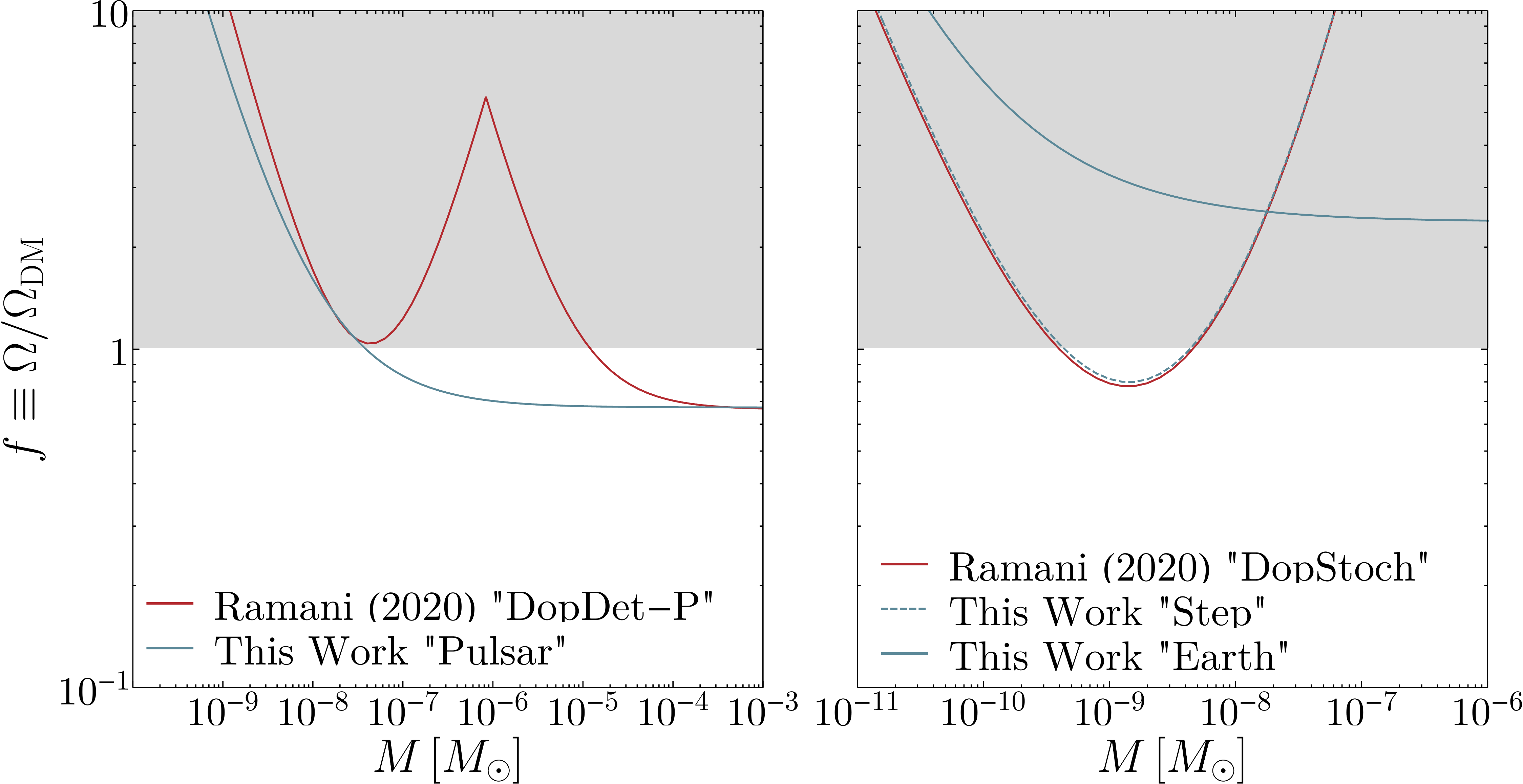}
    \caption{Comparison of the PBH constraints between this work and~\cite{Ramani:2020hdo}. The left panel compares the pulsar term results while the right panel compares the Earth term results. The meanings of the labels are described in the main text. The pulsar parameters used here are $N_p=200$, $\Delta t=2\wk$, $t_{\mathrm{rms}}=50\ns$ and $T=20\yr$. The SNR thresholds are set to ${\rm SNR}=4$ for the pulsar term and ${\rm SNR}=2$ for the earth term for consistency with~\cite{Ramani:2020hdo}.}
    \label{fig:may_comparison}
\end{figure}

The results derived here are more complete and subsume the previous works~\cite{Dror:2019twh, Ramani:2020hdo}. To illustrate the differences with the previous results, in Fig.~\ref{fig:may_comparison} we compare the constraints on the fraction of DM in PBHs, $f = \Omega_\text{PBH}/\Omega_{DM}$. The curves labelled ``DopDet-P" in the left panel and ``DopStoch" in the right panel are taken from~\cite{Ramani:2020hdo}, which are analogous to the ``pulsar term" and ``Earth term" analysis presented in this work. The main difference between the more recent work, Ref.~\cite{Ramani:2020hdo}, and this analysis is how the signal is calculated. Previously the signal $h_I(t)$ and its autocorrelator were computed using analytic approximations, and the constraints were cut off when these were no longer good expressions. Here we explicitly generate the residual signal $h_I$ which avoids the need of analytic simplifications. More specifically, for the ``DopDet-P" curve in~\cite{Ramani:2020hdo} the signal $\delta\phi(t)$ was approximated by the leading order term in the power series in the small (dynamic) and large (static) $\tau/T$ limits. By contrast, this work does not use the aforementioned approximations. This allows the constraint to smoothly interpolate between the two regimes and asymptote to the ``DopDet-P" curves in both limits.
Similarly, the ``DopStoch" curve was obtained using an approximate form of the correlation function $R(t,t')$, which was derived analytically from the step function approximation of $\delta\phi(t)$. As shown in Fig.~\ref{fig:may_comparison}, the constraint has a non-negligible deviation from this work, which is an indication that the approximations used in~\cite{Ramani:2020hdo} is overly optimistic near the peak of the constraint. For an explicit comparison we ran the MC using the same approximate form of $\delta\phi(t)$ in~\cite{Ramani:2020hdo}, keeping only subhalos with impact parameter satisfying $b<\bar{v}T$. The resulting constraints are labelled ``Step" in the right hand panel of Fig.~\ref{fig:may_comparison} and we see good agreement with the result in~\cite{Ramani:2020hdo}.

\section{Comparison between analytic and numerical HMF}
\label{app:PS_comparison}

\begin{figure}[t!]
\begin{center}
\includegraphics[width=0.9\textwidth]{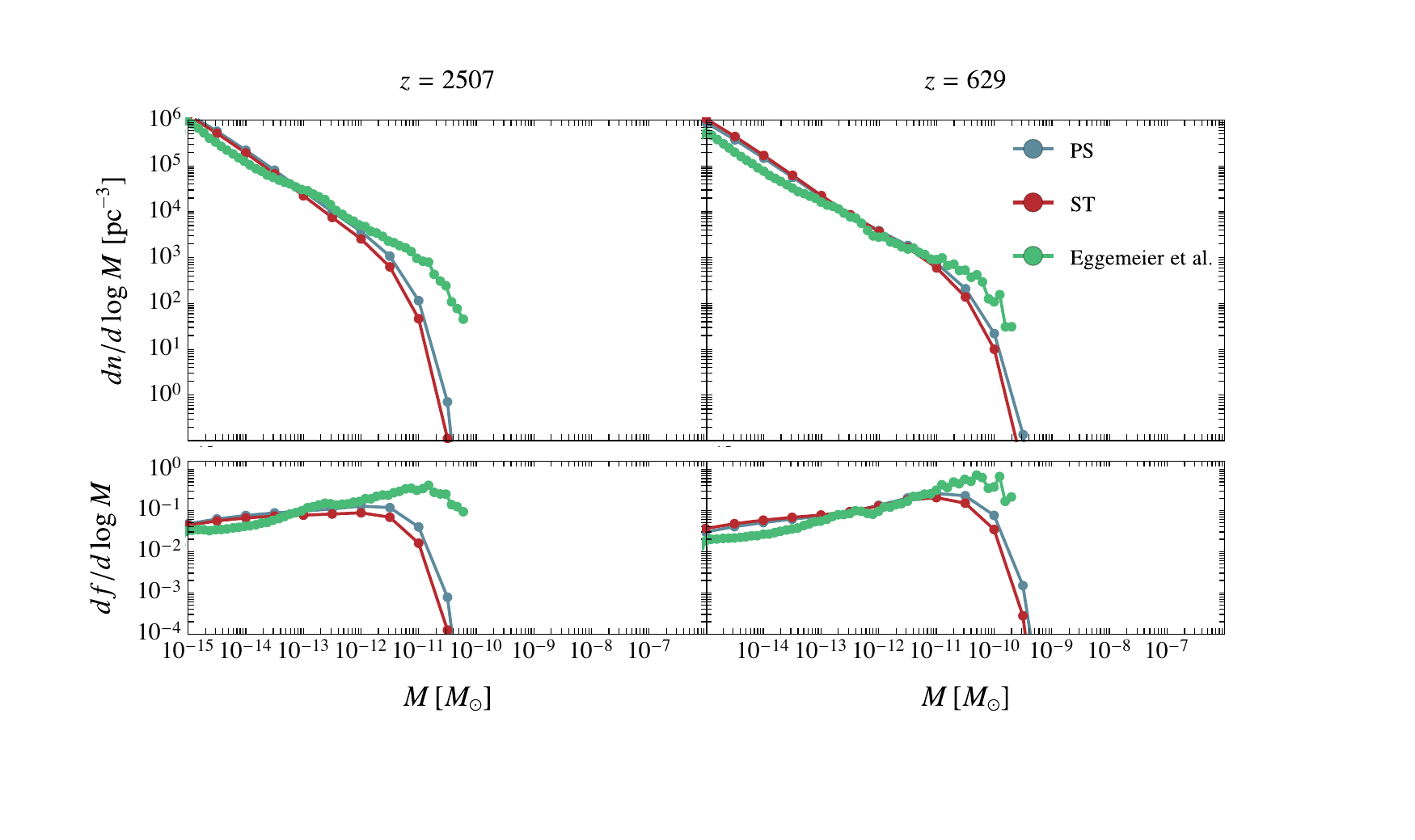}\\
\vspace{1cm}
\includegraphics[width=0.9\textwidth]{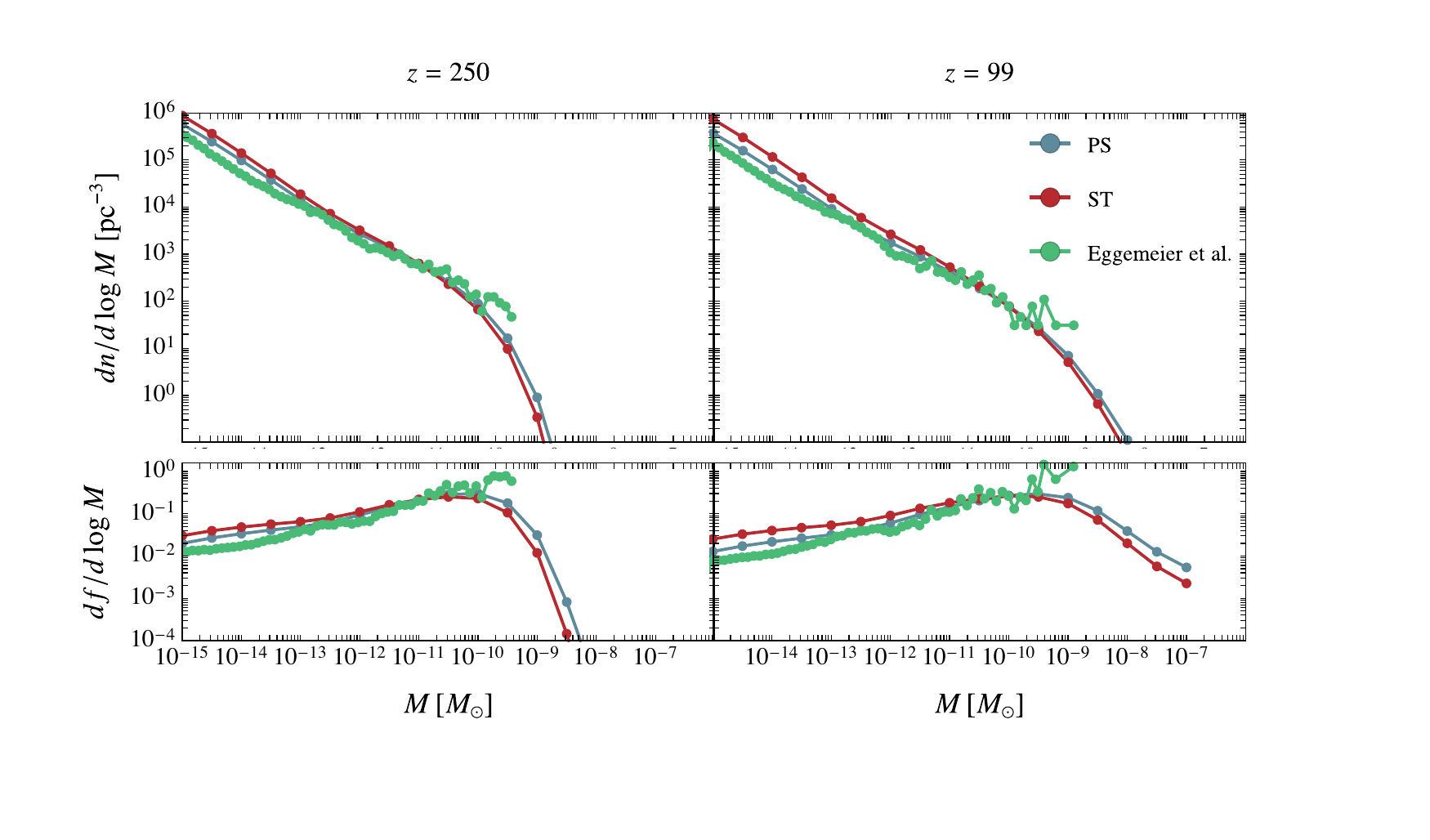}
\caption{Comparison between the HMF predicted by PS (blue) and the results of the numerical simulations (green) at four different redshifts. For comparison we also show the predictions of another analytical prescription (red) developed by Sheth \& Tormen.}
\label{fig:PScomp}
\end{center}
\end{figure}

In this appendix we compare the numerically derived HMF for axion models where the PQ symmetry is broken after inflation \cite{Eggemeier:2019khm}, against some commonly used analytic prescriptions. Specifically, we compare the numerical result against the analytic predictions of the Press-Schechter \cite{Press:1973iz}, and Sheth-Tormen \cite{Sheth:1999mn} formalisms. The first one has been described in the main text, for convenience of reference we report here the differential halo mass fraction predicted by the PS formalism:
\begin{equation}
\frac{df_{\rm PS}(M,z)}{d\ln M}=\sqrt{\frac{2}{\pi}}\nu(M,z)\exp\left(-\frac{\nu^2(M,z)}{2}\right)\frac{d\ln\sigma(M,z)}{d\ln M}\,,
\end{equation}
where $\nu(M,z)\equiv\delta_c/\sigma(M,z)$. The Sheth-Tormen formalism instead predict a differential halo mass fraction given by 
\begin{equation}
\frac{df_{\rm ST}(M,z)}{d\ln M}=A \sqrt{\frac{a}{\pi}}\nu(M,z)\left[1+\left(\frac{1}{a\,\nu^2(M,z)}\right)^p\right]\exp\left(-\frac{a\,\nu^2(M,z)}{2}\right)\frac{d\ln\sigma(M,z)}{d\ln M}\,,
\end{equation}
where the choice of parameters $A=0.3222$, $a=707$, and $p=0.3$ has been showed to best reproduce the numerical results (at least for the the case of $\Lambda$CDM). 

The results of the comparison, at four different redshifts,  are shown in \Fig{fig:PScomp}. The agreement between numerical and analytic results is quite good, with the exception of the high mass regime where simulations start to lose sensitivity. 

\section{SNR statistical significance}\label{app:snr_significance}
In this section we discuss the statistical significance associated with a given observed value of the SNR, more specifically, we derive its $p$-value (\emph{i.e} the probability that an SNR at least as large as the one observed is generated in presence of noise only). 

In absence of a signal, the pulsar SNR given in \eq{eq:s_snr_pulsar} can be written as
\begin{equation}
{\rm SNR}=\frac{\left|\displaystyle\int dt\,n(t)\right|}{\left(\displaystyle\int dt dt'\langle n(t)n(t')\rangle\right)^{1/2}}=\frac{1}{\sqrt{T \widetilde N}}\left|\int dt\,n(t)\right|\,.
\end{equation}
where we have drop the expectation value in the numerator because we want to study the full statistical distribution of the SNR, and ignored the filter functions for simplicity. Assuming that the noise is Gaussian, it is evident from the previous equation that the SNR in absence of a signal is distributed as the absolute value of a Gaussian variable with zero mean and unit standard deviation. From this follow that, given an array of $N_P$ pulsars, the $p$-value for the pulsar term is 
\begin{equation}
p({\rm SNR})=1-\left[{\rm erf}\left(\frac{{\rm SNR}}{\sqrt2}\right)^{N_P}\right]\,.
\end{equation}
   
Assuming pulsar independent noise, the Earth SNR in absence of a signal takes the form 
\begin{equation}\label{eq:snr_earth_noise}
{\rm SNR}=\frac{\left|\sum\limits_{I,J}\displaystyle\int dtdt'n_I(t)n_J(t')\right|}{\widetilde N\sqrt{2 N_P(N_P-1)}}
\end{equation}
where, as before, we have dropped the filter functions and the expectation value in the numerator. We now define 
\begin{equation}
\rho\equiv\frac{\sum_{I}\int dt\,n_I(t)}{\sqrt{\widetilde N N_P T}}
\end{equation}
which, by construction, is a Gaussian variable with zero mean and unit standard deviation. In terms of $\rho$, Eq.~\eqref{eq:snr_earth_noise} can be rewritten as 
\begin{equation}
{\rm SNR}=\frac{N_P}{\sqrt{2N_P(N_P-1)}}\rho^2\,,
\end{equation}
from which conclude that, in absence of signal, the Earth SNR follows a rescaled $\chi$-squared distribution. Therefore, the $p$-value for the Earth term is given by 
\begin{equation}
p({\rm SNR})=1-P(1/2,\frac{\sqrt{2N_P(N_P-1)}}{N_P}\,{\rm SNR})
\end{equation}
where $P(s,t)$ is the regularized gamma function. 

Finally, when showing our final results in the main text, we express $p$-values in terms of standard deviations by using the relation
\begin{equation}
\sigma_{\rm significance}(p)=\sqrt{2}{\rm erf}^{-1}(1-2p)\,.
\end{equation}
As an example, an ${\rm SNR}=4$ corresponds (both for the pulsar and Earth term) to a p-value of $p\sim0.01$, and a signal significance of $\sigma_{\rm significance}\sim 2$. 

\bibliography{bibliography}

\end{document}